\documentclass[12pt,a4paper]{article}

\pdfoutput=1 
\usepackage[utf8]{inputenc}
\usepackage{fullpage}
\usepackage{cite}
\usepackage{subcaption}
\usepackage[english]{babel}
\usepackage{amsmath}
\usepackage{physics}
\usepackage{amsthm}
\usepackage{amsfonts}
\usepackage{amssymb}
\usepackage{bbold}
\usepackage{mathrsfs}
\usepackage{mathbbol}
\usepackage[toc,page]{appendix}
\usepackage{graphicx}
\usepackage{bbding}
\usepackage{esvect}
\usepackage{commath}
\usepackage{enumitem}
\usepackage[affil-it]{authblk}
\usepackage{siunitx}
\usepackage{braket}
\usepackage{phaistos}
\usepackage{multicol}
\usepackage{overpic}
\usepackage{comment}
\usepackage[colorlinks=true
  ,urlcolor=blue
  ,anchorcolor=blue
  ,citecolor=blue
  ,filecolor=blue
  ,linkcolor=blue
  ,menucolor=blue
  ,pagecolor=blue
  ,linktocpage=true
  ,pdfproducer=medialab
  ,pdfa=true
]{hyperref}
\usepackage{xcolor}

\usepackage{jheppub}
\usepackage{tikz}

\usepackage[most]{tcolorbox}

\definecolor{light-gray}{gray}{0.95}
\tcbset{textmarker/.style={%
        enhanced,
        parbox=false,boxrule=0mm,boxsep=0mm,arc=0mm,
        outer arc=0mm,left=6mm,right=3mm,top=7pt,bottom=7pt,
        toptitle=1mm,bottomtitle=1mm,oversize}}
    \newtcolorbox{inputBox}{textmarker,
        borderline west={6pt}{0pt}{black},
        colback=black!2!white}
    \newtcolorbox{outputBox}{textmarker,
        borderline west={6pt}{0pt}{black},
        colback=black!8!light-gray}

\newcommand{\inBox}[1]{\begin{inputBox} \textbf{Input:} #1 \end{inputBox}}
\newcommand{\outBox}[1]{\begin{outputBox} \textbf{Output:} #1 \end{outputBox}}

\usepackage{bbm}

\newcommand{\Hil}{\mathcal{H}}
\newcommand{\Mod}[1]{\,\mathrm{mod}\,#1}
\newcommand{\Stab}[1]{\,\textnormal{Stab}\,#1}
\newcommand{\HC}{(HC)_{1,2}}

\def\ket#1{{|{#1}\rangle}} 

\title{Clifford Orbits from Cayley Graph Quotients}

\author[a]{Cynthia Keeler,}
\author[a]{William Munizzi,}
\author[b,c]{and Jason Pollack}

\affiliation[a]{Department of Physics, Arizona State University,
Tempe, AZ 85281, USA}
\affiliation[b]{Quantum Information Center, Department of Computer Science, The University of Texas at Austin,
2317 Speedway, Austin, TX 78712, USA}
\affiliation[c]{Department of Electrical Engineering and Computer Science, Syracuse University, NY 13210, USA}

\emailAdd{keelerc@asu.edu}
\emailAdd{wmunizzi@asu.edu}
\emailAdd{jasonpollack@gmail.com}

\abstract{We describe the structure of the $n$-qubit Clifford group $\mathcal{C}_n$ via Cayley graphs, whose vertices represent group elements and edges represent generators. In order to obtain the action of Clifford gates on a given quantum state, we introduce a quotient procedure. Quotienting the Cayley graph by the stabilizer subgroup of a state gives a reduced graph which depicts the state's Clifford orbit. Using this protocol for $\mathcal{C}_2$, we reproduce and generalize the reachability graphs introduced in \cite{Keeler2022}. Since the procedure is state-independent, we extend our study to non-stabilizer states, including the W and Dicke states. Our new construction provides a more precise understanding of state evolution under Clifford circuit action.}

\begin{document} 
\maketitle
\flushbottom

\section{Introduction}\label{Intro}

How can we track the evolution of information about a quantum state? 
In the general case, where we'd like to obtain the outcome of arbitrary measurements under continuous time evolution, we can do no better than to work with the state itself. 
For an $n$-qubit pure state $\ket{\psi}\in \Hil \cong \mathbb{C}^{2n}$, tracking the state requires knowing all of its overlaps $\braket{a_i|\psi}$ with a given orthonormal basis $\{\ket{a_i}\}$: that is, $4^n-2$ real parameters, accounting for normalization and the unobservability of global phase.
We can do better, however, by restricting which information we want to keep track of, or restricting how the state might evolve, or starting with a special initial state:
\begin{itemize}
    \item We might only care about some particular properties of the state. 
    If we only want to predict the outcome of measurements on $k<n$ of the qubits, we can trace out the remaining $n-k$ qubits and work with the reduced state $\rho_{\{n-k\}}$, which requires only $4^k-1$ real parameters. 
    If we want to understand the entanglement properties of the state, we can collect together the von Neumann entropies of each independent reduced density matrix: $2^{n-1}-1$ real parameters.
    \item We might want to evolve the state through a quantum circuit with a limited set of possible unitaries that can be applied, such as the Clifford gates: Hadamard, phase, and CNOT. It might be that the multiplicative group spanned by the gate set is sufficient to approximate any unitary acting on $\mathbb{C}^{2n}$, up to arbitrary accuracy, in which case the gate set is universal. 
    But we might instead find that the group acts on a smaller Hilbert space, for example if every gate in the gate set conserves some charge. 
    Or, as is the case for the Clifford gates \cite{Calderbank:1996hm,Gottesman:1997qd}, the generated group might be \emph{finite}, in which case, for a given initial state, there are only a discrete set of possible states which can be reached.
    \item We might have a special state that admits a reduced description. 
    If we know our state has decohered, we can describe it by a classical ensemble of pointer states, and classical observables are independent of the relative phase between branches, requiring only $2^n-1$ real parameters. 
    Or, if we know our state is an eigenstate of some specified observable, or a simultaneous eigenstate of a group of observables, we can obtain a compact description.
    For example, the Pauli group on $n$ qubits comprises the $2\cdot4^n$ (signed) Pauli strings. 
    All $n$-qubit states stabilize, i.e.\ are unit eigenvectors of, the identity operator $\mathbb{1}$ (and none stabilize $-\mathbb{1}$). 
    But only a discrete set of states stabilize any additional Pauli strings: a special set of states that can be specified by discrete rather than continuous information.
\end{itemize}

One of the most famous results in the theory of quantum computation concerns a computational setting where we allow ourselves such simplifications. Quantum circuits which take an initial stabilizer state\footnote{This paper, unfortunately, will have to deal with two meanings of the word `stabilizer': the group-theoretic meaning, where a state stabilizes a group element if the group element acts trivially on the state, and the quantum-information-theoretic, where it is standard to refer to those $n$-qubit states which stabilize $2^n$ elements of the $n$-qubit Pauli group simply as ``stabilizer states''. 
We will have cause in this paper to refer to \emph{both} the traditional stabilizer states \emph{and} states which stabilize elements of other groups, most notably the $n$-qubit Clifford group and its subgroups.}, a simultaneous unit eigenvector of $2^n$ Pauli strings, to any other stabilizer state can be represented as ``Clifford circuits'', which contain only Clifford gate applications. 
Unlike circuits made from a universal quantum gate set, Clifford circuits are efficiently classically simulable \cite{Gottesman:1997zz,Gottesman:1998hu,aaronson2004improved}. 

In this paper, we exploit the finiteness of the Clifford group to reinterpret Clifford circuits graphically. 
Our key tool is a group-theoretic notion: the Cayley graph \cite{10.2307/2369306}, which, given a choice of generators, graphically encodes the structure of the group. We're interested in studying what states can be reached if, instead of acting with \emph{arbitrary} Clifford circuits, we restrict to only a subset of the possible Clifford gates. 
In particular, following our previous paper \cite{Keeler2022}, we'd like to understand how the entanglement entropy evolves as we act on a state with a Clifford circuit. 
Because all of the entanglement created in this way is bipartite---the result of a $CNOT$ gate action---it suffices to consider entangling operations acting on only $2$ of the $n$ qubits. 

We are thus led to consider the actions of the $2$-qubit Clifford group on $n\ge 2$-qubit states, which might themselves be stabilizer states, or might be more general.
We previously considered a version of this problem in \cite{Keeler2022}, developing ``reachability graphs'' in which each vertex was a stabilizer state and each edge a Clifford gate, and ``restricted graphs'' in which only certain types of edges were allowed.
We found that the complicated graphs encoding the action of Clifford circuits on stabilizer states decomposed into highly structured subgraphs.

In this paper, using the technology of Cayley graphs, we are able to reproduce and generalize our previous results. 
Rather than working with the action of Clifford gates on states, we are instead led to work more directly with the abstract group elements themselves. We can then recover the action on states via a quotienting procedure. 
By working group-theoretically, we can easily understand the full diversity of subgraph structures that arise, as well as extend our results to the action of Clifford circuits on other states --- for example, states which stabilize a non-maximal number of Pauli group elements --- which allows for new structures to arise. 
Along the way, we will gain a better understanding of the Clifford group itself, deriving a formal presentation for the group, as well as data on its subgroup structure. While presentations and descriptions of the Clifford group exist in the literature \cite{Selinger2013,Farinholt_2014}, our reformulation gives insight into the circumstances in which seemingly entangling gate operations fail to ultimately produce entanglement. 

In \cite{Keeler:2023shl} we use the relations of our presentation to examine and bound the dynamics of entanglement entropy. Because holographic states, which have a classical geometric description, live inside the stabilizer entropy cone \cite{Bao:2015bfa}, tracking the evolution of entanglement can give us insight into the operations which move quantum states into and out of the holographic cone. 

\subsection{Summary of Results}

We previously introduced, in \cite{Keeler2022}, \emph{reachability graphs}, in which each vertex is an $n$-qubit pure quantum state, typically a stabilizer state, and each (directed) edge is a Clifford gate taking the state at the initial vertex to the state at the final vertex, as well as \emph{restricted graphs}, in which only some subset of the Clifford gates, typically $\{H_1,\, H_2,\, C_{1,2},\, C_{2,1}\}$, is allowed. (Here and throughout we abbreviate $CNOT_{i,j}$ as $C_{i,j}$.)
Reachability graphs graphically encode the result of performing Clifford circuits on a given set of states; restricted graphs give a more refined picture which is often more useful for understanding entropic evolution.  

The main task accomplished in this paper is the reinterpretation of reachability graphs as certain quotients of a group-theoretic object, \emph{the Cayley graph}, a directed graph that encodes the structure of a group by identifying a vertex for every group element and a set of edges for each group generator.%
\footnote{Because the Cayley graph depends on a choice of generators, there are many different Cayley graphs which each correspond to a given group. Each of these graphs has an isomorphic set of vertices, namely one vertex for every group element, but in general inequivalent edges. For example, $\{H_1,\, H_2,\, C_{1,2},\, C_{2,1}\}$ and $\{H_1,\, H_2,\, C_{1,2}\}$ both generate the two-qubit Clifford group; as seen in Table \ref{tab:OrbitLengthCliffordSubgroupNoRelations}, the corresponding Cayley graphs have the same number of vertices, $2304$, but different properties such as graph diameter.} %
Since finite groups have finite Cayley graphs, we can use group cosets to construct quotient spaces on its Cayley graph.

The general protocol for quotienting a Cayley graph to yield a graph isomorphic to a reachability graph is:
\begin{enumerate}
    \item For a group $G$ and chosen state $\ket{\Psi}$, we first identify the stabilizer subgroup of $\ket{\Psi}$ in $G$, denoted $\Stab_G(\ket{\Psi})$.
    \item Since $\Stab_G(\ket{\Psi})$ is a subgroup of $G$, all equivalence classes of the left coset space $G/\Stab_G(\ket{\Psi})$ can be generated by taking $h\cdot\Stab_G(\ket{\Psi})$ for all $h \in G$.
    \item Each equivalence class of $G/\Stab_G(\ket{\Psi})$ is assigned a vertex, and each vertex is connected by the generator which maps each element of one equivalence class to exactly one element of the other equivalence class.
\end{enumerate}

\begin{figure}[h]
    \centering
    \includegraphics[width=14.5cm]{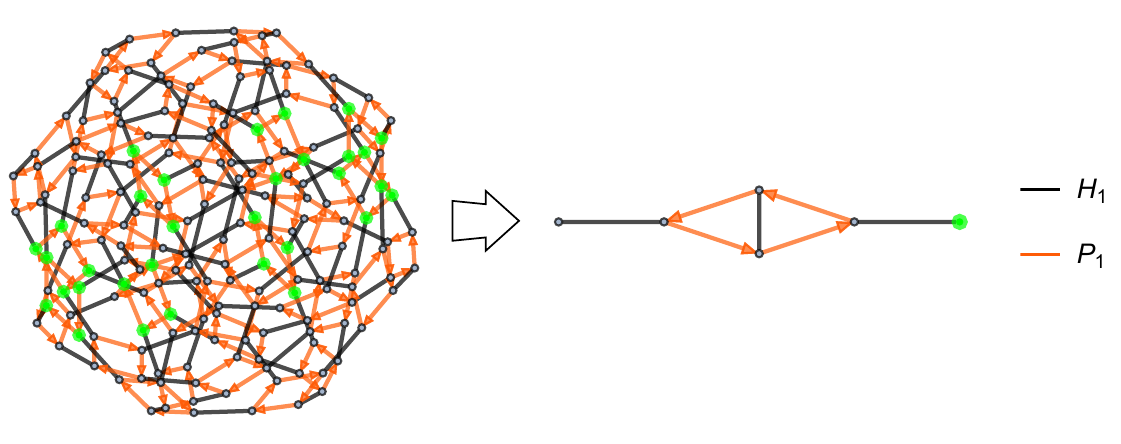}
    \caption{There are $32$ elements of the single-qubit Clifford group $\mathcal{C}_1$ which form the equivalence class of stabilizers for $\ket{0}$, denoted $\Stab_{\mathcal{C}_1}(\ket{0})$. We first build the quotient group $\bar{\mathcal{C}_1} \equiv \mathcal{C}_1/\langle \omega\rangle$ by modding out global phase, then build the left coset space $\bar{\mathcal{C}_1}/\Stab_{\bar{\mathcal{C}_1}}(\ket{0})$ to identify equivalent vertices in the $\mathcal{C}_1$ Cayley graph. This process yields a quotient space of the Cayley graph which is isomorphic to the one-qubit stabilizer reachability graph.}
    \label{SummaryImagePaper}
\end{figure}

This procedure takes as input a choice of group and a choice of state.
To recover the reachability graphs, we do not take $G$ to be the Clifford group itself, because the group contains elements which act as an (unobservable) global phase. 
Instead, we first quotient the Clifford group (or $\HC$, the group generated by Hadamard and CNOT acting on the first two qubits) by such elements, resulting in a smaller group whose elements are isomorphic to equivalence classes of the original group; only then do we specify a state. 
Figure \ref{SummaryImagePaper} illustrates this procedure, starting with the one-qubit Clifford group and the chosen state $\ket{0}$ and producing a quotiented graph isomorphic to the one-qubit stabilizer reachability graph. 
We could have started with any of the six one-qubit stabilizer states and gotten the same result, but the precise mapping of initial group elements to vertices of the final Cayley graph is state-dependent.

%

In the remainder of the paper, we build an understanding of the Clifford group detailed enough to construct its Cayley graph and those of its subgroups. Then, with that task accomplished, we display the diversity of structures which ensue from applying the procedure to various quantum states. 
In Section \ref{sec:reminder}, we recall the Pauli group, Clifford group, stabilizer states, and the reachability graphs defined in our previous paper. 
In Section \ref{OneQubitSection}, we begin with the simple case of the one-qubit Clifford group $\mathcal{C}_1$, writing down its presentation and displaying its Cayley graph. 
We accomplish the same task for the two-qubit Clifford group $\mathcal{C}_2$ in Section \ref{TwoQubitSection}, where both of these steps are considerably more complicated. 
We also discuss those subgroups of $\mathcal{C}_2$ generated by a proper subset of the Clifford gates, presenting in Table \ref{tab:OrbitLengthCliffordSubgroupNoRelations} a comprehensive list of these subgroups and their properties, which might be of independent interest, as well as discussing a number of the groups in more detail. 
With the needed group-theoretic data obtained, we proceed in Section \ref{ReachabilityGraphsFromCayleyGraphs} to defining in detail the quotienting procedure summarized above and applying it to various groups and states of interest. 
We summarize and discuss further directions in Section \ref{sec:discussion}. 
Appendices \ref{ExtendedTableAppendix}-- \ref{AdditionalGraphsAppendix} provide further details of our derivations. All Mathematica data and packages are publicly available \cite{githubStab,githubCayley}, and code for generating colorblind-accessible graphs is available upon request.

\section{Reminder: Clifford Group and Reachability Graphs}\label{sec:reminder}

We begin with a brief review of background material discussed throughout this paper. Much of this review was covered more extensively in Section 2 of \cite{Keeler2022}, and we invite the interested reader to consult it for additional details. Likewise, for a more pedagogical reference on the group-theoretic concepts we recommend e.g.\ \cite{Alperin1995}.

\subsection{The Clifford Group}

The Pauli matrices are a set of unitary and Hermitian matrices with $\pm 1$ eigenvalues, defined
\begin{equation}
    \mathbb{1}\equiv\begin{bmatrix}1&0\\0&1\end{bmatrix}, \,\, \sigma_X\equiv\begin{bmatrix}0&1\\1&0\end{bmatrix}, \,\,
    \sigma_Y\equiv\begin{bmatrix}0&-i\\i&0\end{bmatrix}, \,\,
    \sigma_Z\equiv\begin{bmatrix}1&0\\0&-1\end{bmatrix}.
\end{equation}
These matrices act as operators on a Hilbert space $\mathbb{C}^2$ in the fixed measurement basis $\{\ket{0}, \ket{1}\}$. The Pauli operators $\{\sigma_x, \sigma_y, \sigma_z \}$ generate the algebra of all linear operators on $\mathbb{C}^2$, and define a 16-element multiplicative matrix group, the one-qubit Pauli group:
\begin{equation}
    \Pi_1\equiv\langle\sigma_X,\sigma_Y,\sigma_Z\rangle,
\end{equation} 

The set of unitary matrices that normalize the Pauli group is known as the (one-qubit) Clifford group,
\begin{equation}
    \mathcal{C}_1\equiv\left\{ U\in L(\mathbb{C}^2) \: | \: 
    UgU^\dagger\: \forall g \in \Pi_1\right\}.
\end{equation}
Elements of $\mathcal{C}_1$ act as automorphisms on $\Pi_1$ via conjugation by $U$. The single-qubit Clifford group $\mathcal{C}_1$ is generated by the Hadamard and phase quantum gates, defined in a matrix representation as
\begin{equation}\label{HAndPMatrices}
    H \equiv \frac{1}{\sqrt{2}}\begin{bmatrix} 1 & 1 \\
    1 & -1\\
    \end{bmatrix}, \qquad P \equiv \begin{bmatrix} 1 & 0 \\
    0 & i\\
    \end{bmatrix}.
\end{equation}

We can extend the action of the Pauli group and Clifford groups to multiple qubits by composing strings of operators. These ``Pauli strings'' generalize local Pauli group action to a selected qubit in an $n$-qubit system, e.g. the operator which acts with $\sigma_Z$ on only the $k^{th}$ qubit can be written
\begin{equation}\label{PauliString}
    I^1\otimes\ldots\otimes I^{k-1} \otimes \sigma_Z^k \otimes I^{k+1} \otimes \ldots \otimes I^n.
\end{equation}
The weight of a Pauli string refers to the number of non-identity insertions in its tensor product representation. Eq. \eqref{PauliString} shows a Pauli string of weight one, and the set of all weight-one Pauli strings is sufficient to generate%
 \footnote{Constructing Pauli strings requires both a factorization of the $2^n$ dimensional Hilbert space in some fixed basis, as well as a chosen ordering of these factors $\{1,...,n\}$. We choose the ordering $\ket{a_1\ldots a_n}\equiv \ket{a_1}_1\otimes\ldots\ket{a_n}_n$. Elements of the $n$-qubit Pauli group are independent of any ordering choice, as are elements of the $n$-qubit Clifford group; however, the matrix representation of specific gates will depend on this order. We often consider groups which act on an $\ell$-qubit subsystem of an $n$-qubit state, fixed by a choice of ordered indices.} %
the $n$-qubit Pauli group $\Pi_n$.

The construction of the Clifford group can likewise be extended to $n>1$ qubits by adding CNOT gates to the generating set. The CNOT gate $C_{i,j}$ acts bi-locally on two qubits, performing a NOT operation on the $j^{th}$ qubit depending on the state of the $i^{th}$ qubit. In our matrix representation, we write $C_{i,j}$ as
\begin{equation}
    C_{i,j} = 
    \begin{bmatrix} 
    1 & 0 & 0 & 0 \\
    0 & 1 & 0 & 0 \\
    0 & 0 & 0 & 1 \\
    0 & 0 & 1 & 0 \\
    \end{bmatrix},
\end{equation}
where $i$ denotes the control bit and $j$ the target bit. We emphasize the fact that $C_{i,j} \neq C_{j,i}$. The group $\mathcal{C}_n$ is then
\begin{equation}
    \mathcal{C}_n \equiv \langle H_1,...,\,H_n,\,P_1,...,P_n,\,C_{1,2},\,C_{2,1},...,\,C_{n-1,n},\,C_{n,n-1}\rangle.
\end{equation}
We use a similar scheme when representing local gates, where the index denotes the qubit being acted on, e.g. $H_1$.

In this paper, we construct a presentation for the groups $\mathcal{C}_1$ and  $\mathcal{C}_2$, and analyze subgroups which are built by restricting the generating set. A presentation specifies a group by choosing a set of generators and fixing a set of relations among those generators. Elements of the group are then constructed by composing generators using the group operation, subject to the constraints set by the relations.

Every element of a multiplicative group can be written as a product of generators, known as a word. Words which independently equate to the same group element can be transformed into each other using relations in the presentation. In this way, unique words composed of Clifford group generators correspond to different constructible stabilizer circuits. We will present our set of relations as equalities between words built from Clifford generators.

\subsection{Stabilizer Formalism and Reachability Graphs}

For a group $G\subset L(\Hil)$ acting on a Hilbert space we define the stabilizer subgroup $\Stab_G(\ket{\Psi}) \leq G$, for some $\ket{\Psi} \in \Hil$, as the set of elements that leave $\ket{\Psi}$ unchanged,
\begin{equation}\label{StabilizerGroupDefinition}
    \Stab_G(\ket{\Psi}) \equiv \{g\in G\: | \:g\ket{\Psi}=\ket{\Psi}\}.
\end{equation}
That is, $\Stab_G(\ket{\Psi})$ contains only the elements of $G$ for which $\ket{\Psi}$ is an eigenvector with eigenvalue $+1$.

To make further use of this group-theoretic concept, we invoke two important theorems \cite{Alperin1995}. First, given a finite group $G$ and subgroup $H \leq G$, Lagrange's theorem states that the order of $H$ gives an integer partition of $G$, that is
\begin{equation}
    |G| = [G:H] \cdot |H|, \quad \forall \,H \leq G,
\end{equation}
where $[G:H]$ denotes the index of $H$ in $G$. Subsequently the Orbit-Stabilizer theorem says that, when considering the action of $G$ on a set $X$ and $H=\Stab_G(x)$, the orbit of $x \in X$ under $G$ has size
\begin{equation}\label{OrbitStabilizerTheorem}
    [G\cdot x] = [G:H] = \frac{|G|}{|H|}, \quad \forall \, x \in X.
\end{equation}

Considering the action of $\Pi_n$ on $\Hil$, it is clear that all states are trivially stabilized by $\mathbb{1}$. Certain states, however, are stabilized by additional elements of $\Pi_n$. The $n$-qubit ``stabilizer states'' are those which are stabilized by a subgroup of $\Pi_n$ of size $2^n$, the largest allowed size for an $n$-qubit state \cite{aaronson2004improved,garcia2017geometry}. In general, the set of $n$-qubit stabilizer states contains
\begin{equation}
    |S_n| = 2^n \prod_{k=0}^{n-1}(2^{n-k}+1)
\end{equation}
states \cite{doi:10.1063/1.4818950}. The set $S_n$ can be generated by starting with a state in the measurement basis, typically $\ket{0}^{\otimes n}$, and acting on that state with all elements of $\mathcal{C}_n$. In this way $S_n$ is the orbit $\left[\mathcal{C}_n \cdot \ket{0}^n\right]$. 

This method to generate $S_n$, by acting with $\mathcal{C}_n$ on $\ket{0}^{\otimes n}$, lends itself to a natural graph-theoretic description. By assigning a vertex to each state in the orbit $\left[\mathcal{C}_n \cdot \ket{0}^n\right]$, and an edge to every $\mathcal{C}_n$ generator, the evolution of $\ket{0}^{\otimes n}$ through $\Hil$ generates a discrete and finite graph. This structure, introduced in \cite{Keeler2022}, is known as a reachability graph, as we discussed further in Section \ref{Intro} above. When the action of a proper subgroup of $\mathcal{C}_n$ rather than the group itself is considered, we often use the term ``restricted graph'' to describe the reachability graph.

\section{$\mathcal{C}_1$ Presentation and Cayley Graph}\label{OneQubitSection}

In this section we give a presentation for the one-qubit Clifford group $\mathcal{C}_1$ and construct its Cayley graph. We will use this understanding of $\mathcal{C}_1$ to build a presentation for $\mathcal{C}_2$, as well as its subgroups, in Section \ref{TwoQubitSection}. We demonstrate that restricting the set of generators builds subgraphs of the $\mathcal{C}_1$ Cayley graph. We show that quotienting by a global phase reduces $\mathcal{C}_1$ to the symmetric group $S_4$.

 The one-qubit Clifford group $\mathcal{C}_1$ is generated by $\{H_i,P_i\}$, whose matrix representations are given in Eq. \eqref{HAndPMatrices}. Here $i \in \{1,n\}$ is the qubit being acted on in an $n$-qubit system. Relations \ref{HSquared}, \ref{PFourth}, and \ref{HPComm} give a presentation\footnote{All presentations in this paper were verified using the Magma computer algebra system \cite{MR1484478}. Additional details and code can be found in Appendix \ref{MagmaAppendix}.} for $\mathcal{C}_1$,
\begin{align}
   \quad H_i^2 &= \mathbb{1}, \label{HSquared}\\
   \quad P_i^4 &= \mathbb{1}, \label{PFourth}\\ 
  (H_iP_i)^3 &= (P_iH_i)^3 = \omega \label{HPComm}, 
\end{align}
where $\omega^8 = \mathbb{1}$ acts as a global phase\footnote{There exist additional relations involving Clifford gates and $\omega$. Some notable ones which are used in Section \ref{TwoQubitSection} include $(H_iP_jC_{i,j})^6 = (H_iC_{i,j}P_j)^6 = \omega^6$.} for the group. Eqs. \eqref{HSquared}--\eqref{HPComm} can be directly verified by examining the matrix representations in Eq. \eqref{HAndPMatrices}. All elements of $\mathcal{C}_1$ act locally on qubits, and therefore cannot generate or modify entanglement in a physical system.

\begin{figure}[h]
        \centering
        \includegraphics[width=8.4cm]{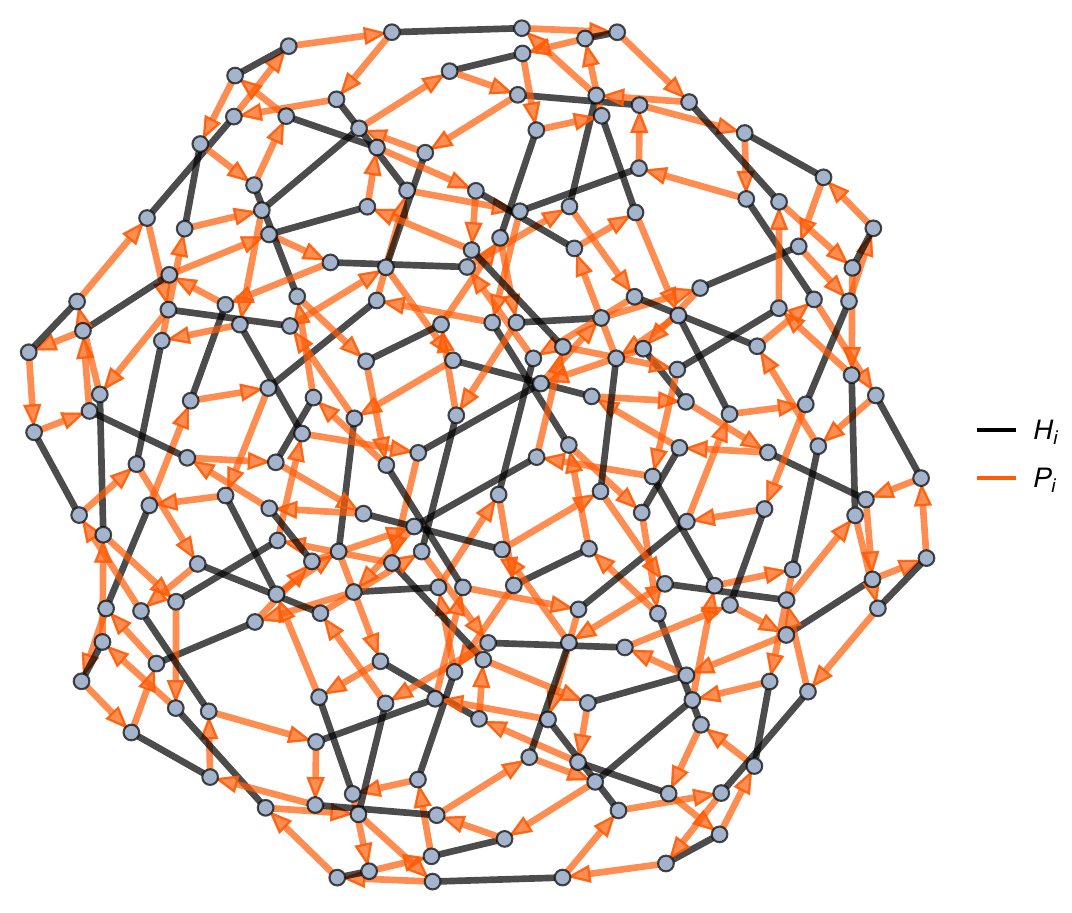}
        \caption{Cayley graph of $\mathcal{C}_1$, with vertices representing group elements and edges representing the generators $H_i$ and $P_i$. We use undirected edges for $H_i$ since $H_i^2 = \mathbb{1}$. The graph has $192$ vertices and $384$ edges, and completely encodes the $\mathcal{C}_1$ group structure.}
    \label{C1CayleyGraph}
    \end{figure}

A \textit{Cayley graph} \cite{10.2307/2369306}, for a group $G$, is built by assigning a vertex to every element in $G$, and an edge for each generator of $G$. The structure of $\mathcal{C}_1$ can be visualized in the Cayley graph shown in Figure \ref{C1CayleyGraph}. Edges of this $\mathcal{C}_1$ Cayley graph represent $H_i$ and $P_i$, while vertices indicate the $192$ unique group elements. Since $H_i^2 = \mathbb{1}$, we use a single undirected edge to represent $H_i$. Directed edges are used to represent $P_i$, as $P_i^2 \neq \mathbb{1}$. In this Cayley graph representation, sequential products of group elements exist as graph paths. Different paths which start and end on the same pair of vertices represent products whose action on the initial element is identical. Loops in the Cayley graph correspond to a sequence of operations which act as the identity.

\begin{figure}[bh]
		\begin{center}
		\begin{overpic}[width=0.35\textwidth]{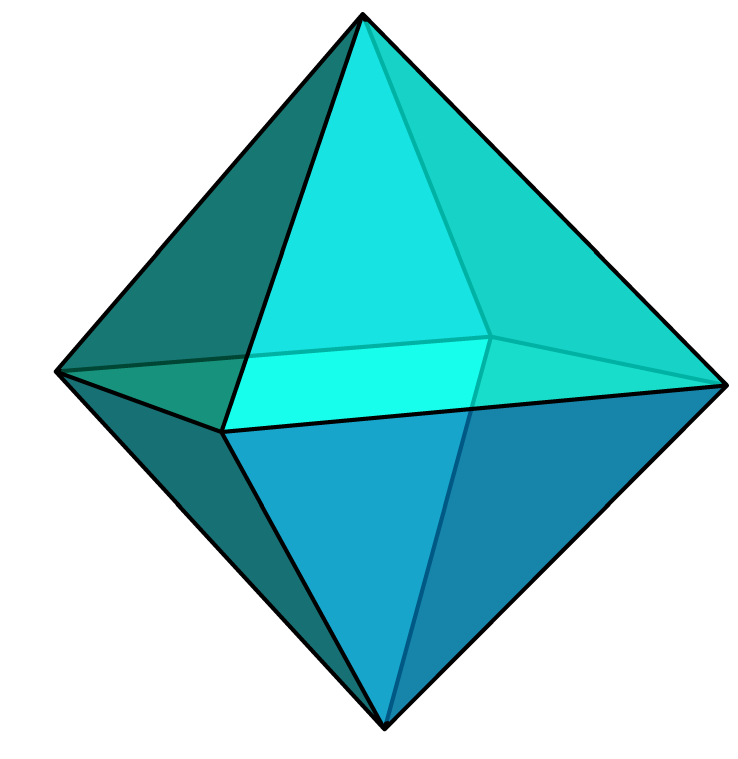}
		\put (44,101) {$\ket{0}$}
		\put (47,-3) {$\ket{1}$}
		\put (-4,50) {$\ket{+}$}
		\put (97,48) {$\ket{-}$}
		\put (23,36.5) {$\ket{i}$}
		\put (66,58) {$\ket{-i}$}
        \end{overpic}
        \caption{The stabilizer octahedron, with $6$ single-qubit stabilizer states at the corners, is often shown embedded in the Bloch sphere. The group $\mathcal{C}_1$, after quotienting by $\omega$, gives the $24$ orientation-preserving maps of this octahedron to itself.}
		\label{StabilizerOctahedron}
	\end{center}
	\end{figure}
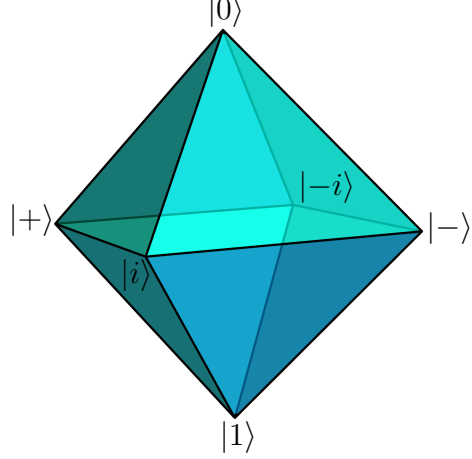

When we ignore global phase, i.e.\ distinguish group elements only up to the factor $\omega$, $\mathcal{C}_1$ reduces to a quotient group with $24$ elements. This quotient group, is isomorphic to $S_4$, the symmetric group of degree $4$; we give its Cayley graph later in the paper, in Figure \ref{C1Quotient}. $S_4$ describes the rotational symmetries of an octahedron, like the well-known stabilizer octahedron shown in Figure \ref{StabilizerOctahedron}. 

\begin{figure}[h]
        \centering
        \includegraphics[width=9cm]{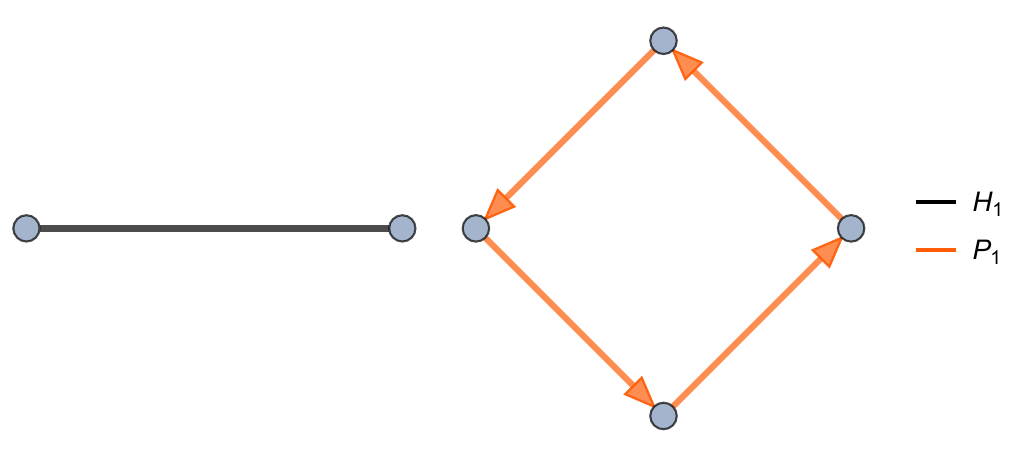}
        \caption{The Cayley graphs for $\mathcal{C}_1$ subgroups $\langle H_i \rangle$ and $\langle P_i \rangle$.}
    \label{HadamardAndPhaseCayleyGraphs}
    \end{figure}

In addition to modding by global phase, we can also construct subgroups of $\mathcal{C}_1$ by restricting our set of generators. The single-generator subgroups $\langle H_i \rangle$ and $\langle P_i \rangle$ are completely described by relations \ref{HSquared} and \ref{PFourth} respectively. The Cayley graphs of $\langle H_i \rangle$ and $\langle P_i \rangle$ are shown in Figure \ref{HadamardAndPhaseCayleyGraphs}.

We gave a presentation for $\mathcal{C}_1$ generated by $H$ and $P$, and introduced $\omega \equiv (H_iP_i)^3$ which acts as a global phase on $\mathcal{C}_1$. We introduced the concept of a Cayley graph, and constructed specific Cayley graphs for $\mathcal{C}_1$ and its single-generator subgroups $\langle H_i \rangle$ and $\langle P_i \rangle$. We described a quotient procedure for groups, and use it to quotient $\mathcal{C}_1$ by $\omega$ to recover $S_4$. Later, we will implement this quotient by $\omega$, and identification up to state stabilizer subgroup, to generate reachability graphs from Cayley graphs. 

\section{$\mathcal{C}_2$ Presentation and Cayley Graphs}\label{TwoQubitSection}

In this section we give a presentation for the two-qubit Clifford group generated by $H, \,P,$ and $CNOT$. This presentation includes a set of operator-level relations, which serve as a set of state-independent constraints on Clifford circuits. We use this presentation to construct all subgroups of $\mathcal{C}_2$ which are generated by subsets of $\{H_i,H_j,P_i,P_j,C_{i,j},C_{j,i}\}$. We give the order of each $\mathcal{C}_2$ subgroup and show how each order is reduced after quotienting the group by $\omega$. For several examples we explicitly build up each element of a subgroup to demonstrate how our relations constrain combinations of Clifford operators.

Every group can be represented by a Cayley graph, which we build for all $\mathcal{C}_2$ subgroups. Since Cayley graphs are state-independent structures, we can use them to study Clifford orbits of arbitrary quantum states. We compute the graph diameter for each $\mathcal{C}_2$ subgroup Cayley graph, both before and after quotienting by $\omega$. We display the Cayley graphs for several example subgroups and highlight group relations that can be visualized as graph paths. In later sections, we will use the quotient procedure outlined here to construct reachability graphs as quotient spaces of Cayley graphs.

\subsection{$\mathcal{C}_2$ Presentation}\label{C2Presentation}

The two-qubit Clifford group $\mathcal{C}_2$ is generated by the set $\{H_i,H_j,P_i,P_j,C_{i,j},C_{j,i}\}$ which consists of the local Hadamard and phase gates, as well as the bi-local CNOT gate. Relations \ref{CSquared}-\ref{ChFourth}, in addition to the $\mathcal{C}_1$ relations \ref{HSquared}-\ref{HPComm}, give a presentation for $\mathcal{C}_2$:
\begin{align}
    C_{i,j}^2 &= \mathbb{1}, \label{CSquared}\\ 
    P_i^{-1}P_jP_i &= P_j, \label{PpComm}\\
    H_i^{-1}H_jH_i &= H_j, \label{HhComm}\\
    P_i^{-1}H_jP_i &= H_j, \label{HpComm}\\
    C_{i,j}H_jC_{i,j}P_jC_{i,j}P_j^3H_j &= P_i, \label{FourGenRelation}\\
    H_iH_jC_{j,i}H_iH_j &= C_{i,j}, \label{CNOTTransform}\\
    (C_{i,j}P_j)^4 &= P_i^2, \label{CpFourth}\\
   \quad C_{i,j}^{-1}C_{j,i}C_{i,j} &= C_{j,i}^{-1}C_{i,j}C_{j,i}, \label{CcComm}\\
    \quad  P_i^{3}C_{i,j}P_i &= C_{i,j}, \label{PCComm}\\
    \quad  (C_{i,j}H_j)^4 &= P_i^2.\label{ChFourth}\\ \nonumber
\end{align}
The relations \ref{CcComm}--\ref{ChFourth}, along with $H_i^2 = \mathbb{1}$ and $P_i^4 = \mathbb{1}$, can be removed to furnish a more minimal presentation%
\footnote{We additionally note that $\mathcal{C}_2$ can be minimally generated from the set $\{H_i, \,H_j, \,P_i, \,C_{j,i}\}$, as can be seen from relations \ref{FourGenRelation} and \ref{CNOTTransform}.} %
using only relations \ref{CSquared}--\ref{CpFourth}. We have nevertheless retained a number of non-minimal relations as they provide insight into the structure of $\mathcal{C}_2$ and will be useful for constructing subgroups in the following subsection. 

Every relation in Eqs. \eqref{CSquared}--\eqref{ChFourth} is a cycle in the Cayley graph of $\mathcal{C}_2$. We especially note relation \ref{ChFourth}, $(C_{i,j}H_j)^4 = P_i^2$, which allows us to build a phase operation using only Hadamard and CNOT. Since $P_i^2$ cannot modify entanglement, neither can the sequence $(C_{i,j}H_j)^4$. This relation is critical for demonstrating entropy bounds on reachability graphs in \cite{Keeler:2023shl}. Relation \eqref{ChFourth} is derived explicitly in \ref{ChFourthEqualsPSquared}. 

While our presentation for $\mathcal{C}_2$ does not depend on the choice of qubits $1$ and $2$, and describes the action of $\mathcal{C}_2$ on an $n$-qubit system, it is not a presentation for $\mathcal{C}_n$ when $n > 2$. A presentation for $\mathcal{C}_n$ requires additional generators for each increase in qubit number. One can, however, generalize our $\mathcal{C}_2$ presentation to a presentation for $\mathcal{C}_3$ by adding only four relations. Each of these four new relations pertain only to Hadamard and CNOT gates, and no new phase gate relations are needed. The additional relations can be found in \cite{Selinger2013}, where alternative presentations%
\footnote{The presentation in \cite{Selinger2013} is given with generators $H_i,\,P_i,$ and $CZ_{i,j}$, and offers a different set of relations. Additionally, $CZ_{i,j} = CZ_{j,i}$ while $C_{i,j} \neq C_{j,i}$.} %
for $\mathcal{C}_1, \,\mathcal{C}_2,$ and $\mathcal{C}_3$ are studied using Clifford circuit normal forms.

\subsection{$\mathcal{C}_2$ Subgroups}

We now give a complete description of all $\mathcal{C}_2$ subgroups built by restricting the generating set. First, we list all such subgroups, as well as their group and Cayley graph properties, in Table \ref{tab:OrbitLengthCliffordSubgroupNoRelations}. We directly construct several subgroups as examples that highlight how our relations constrain strings of Clifford gates at the operator level. We will use these state-independent relations in Section \ref{ReachabilityGraphsFromCayleyGraphs} to build reachability graphs for non-stabilizer quantum states, and further in \cite{Keeler:2023shl} to bound entanglement entropy. 

We can construct subgroups of $\mathcal{C}_2$ by restricting our set of generators to subsets of $\{H_i,H_j,P_i,P_j,C_{i,j},C_{j,i}\}$. One simple case is the subgroup $\mathcal{C}_1$, generated by only $\{H_i,P_i\}$ and discussed in Section \ref{OneQubitSection}. Table \ref{tab:OrbitLengthCliffordSubgroupNoRelations} gives a list of all subgroups constructed in this way. The Table gives the order of each subgroup, as well as the graph diameter (the maximum over all minimum distances between vertices) of the Cayley graph for each subgroup. Since each subgroup below is isomorphic under qubit exchange, we only list one example for each generating set. Subgroups in bold are explicitly constructed in the following text.
\begin{table}[h]
    \centering
    \begin{tabular}{|c||c|c||c|c|}
    \hline
    Generators & Order & Diam. (w/ phase) & Factor & (no phase)\\
    \hline
    \hline
    $\mathbf{\{H_1\}}$ & $2^\dagger$ & 1 & - & - \\
    \hline
    $\mathbf{\{C_{1,2}\}}$ & $2^\dagger$ & 1 & - & - \\
    \hline
    $\mathbf{\{P_1\}}$ & $4$ & 3 & - & - \\
    \hline
    $\mathbf{\{H_1,\,H_2\}}$ & $4$ & 2 & - & - \\
    \hline
    $\mathbf{\{C_{1,2},\,C_{2,1}\}}$ & $6$ & 3 & - & -\\
    \hline
    \{$H_1,\,P_2\}$ & $8^\dagger$ & 4 & - & - \\
    \hline
    \{$P_1,\,C_{1,2}\}$ & $8^\dagger$ & 4 & - & - \\
    \hline
    \{$P_1,\,P_2$\} & $16$ & 6 & - & - \\
    \hline
    \{$H_1,\,C_{2,1}\}$ & $16^\dagger$ & 8 & - & - \\
    \hline
    \{$H_1,\,C_{1,2}\}$ & $16^\dagger$ & 8 & - & - \\
    \hline
    $\mathbf{\{H_1,\,P_2,\,C_{2,1}\}}$ & $32$ & 6 & - & - \\
    \hline
    $\mathbf{\{P_2,\,C_{1,2}\}}$ & $32$ & 8 & - & - \\
    \hline
    \{$P_1,\,P_2,\,C_{2,1}$\} & $64$ & 7 & - & -\\
    \hline
    \{$P_1,\,C_{2,1},\,C_{1,2}$\} & $192$ & 11 & - & - \\
    \hline
    \{$H_1,\,P_1$\} & $192$ & 16 & 8 & 6 \\
    \hline
    \{$H_1,\,H_2,\,P_1$\} & $384$ & 17 & 8 & 7 \\
    \hline
    \{$P_1,\,P_2,\,H_1$\} & $768$ & 19 & 8 & 9 \\
    \hline
    \{$H_1,\,C_{2,1},\,C_{1,2}\}$ & $2304^*$ & 26 & 2 & 15 \\
    \hline
    \{$H_1,\,H_2,\,C_{1,2}\}$ & $2304^*$ & 27 & 2 & 17 \\
    \hline
    $\mathbf{\{H_1,\,H_2,\,C_{1,2},\,C_{2,1}\}}$ & $2304^*$ & 25 & 2 & 15 \\
    \hline
    \{$H_1,\,P_1,\,C_{2,1}\}$ & $3072^*$ & 19 & 8 & 9 \\
    \hline
    \{$H_1,\,P_1,\,C_{1,2}$\} & $3072$ & 19 & 8 & 11 \\
    \hline
    \{$H_1,\,P_1,\,P_2,\,C_{2,1}\}$ & $3072^*$ & 19 & 8 & 9 \\
    \hline
    \{$H_1,\,H_2,\,P_1,\,P_2$\} & $4608$ & 17 & 8 & 12 \\
    \hline
    $\mathbf{\{H_1,\,P_2,\,C_{1,2}\}}$ & $9216$ & 24 & 8 & 13 \\
    \hline
    \{$H_1,\,H_2,\,P_1,\,C_{2,1}\}$ & $92160^*$ & 21 & 8 & 13 \\
    \hline
    \{$H_1,\,H_2,\,P_1,\,C_{1,2}\}$ & $92160^*$ & 21 & 8 & 16 \\
    \hline
    \{$H_1,\,P_1,\,P_2,\,C_{1,2}\}$ & $92160^*$ & 21 & 8 & 14 \\
    \hline
    \{$H_1,\,H_2,\,P_1,\,P_2,\,C_{1,2},\,C_{2,1}\}$ & $92160^*$ & 19 & 8 & 11\\
    \hline
    \end{tabular}
\caption{Subgroups generated by generator subsets are shown in the leftmost column. We give the order of each subgroup and its Cayley graph diameter, both before and after modding by global phase. The third column gives the factor reduction by removing global phase. An asterisk indicates groups with the same elements, and a dagger indicates groups with isomorphic Cayley graphs. Bolded subgroups are explicitly constructed in the text.}
\label{tab:OrbitLengthCliffordSubgroupNoRelations}
\end{table}

Some subgroups have the same order and are isomorphic, e.g. $\langle H_1 \rangle \cong \langle C_{1,2}\rangle$ and $\langle H_1, \,P_1\rangle \cong\langle P_1, \,C_{1,2}\rangle$. Other subgroups have the same order, but are not isomorphic; for example, subgroups $\langle P_1, \,P_2 \rangle,\, \langle H_1, \,C_{1,2} \rangle,$ and $\langle H_1, \,C_{2,1} \rangle$ all have order $16$, but $\langle P_1, \,P_2 \rangle \ncong \langle H_1, \,C_{1,2} \rangle \cong \langle H_1, \,C_{2,1} \rangle$. Even when generated groups are isomorphic, as is the case for subgroups $\langle H_1, \,H_2, \,C_{1,2} \rangle, \,\langle H_1, \,C_{1,2}, \,C_{2,1} \rangle,$ and $\langle H_1, \,H_2, \,C_{1,2}, \,C_{2,1} \rangle$, we emphasize that they may not have isomorphic Cayley graphs, since the Cayley graph depends on not just the group but a choice of generators: here, none of the three descriptions do.
\newpage
Subgroup $\langle H_1, \,H_2, \,C_{1,2} \rangle$, which contains the same elements as $\langle H_1, \,C_{1,2}, \,C_{2,1}\rangle$ and $\langle H_1, \,H_2, \,C_{1,2}, \,C_{2,1}\rangle$, has the Cayley graph of largest diameter. Adding $C_{2,1}$ to the set $\langle H_1, \,H_2, \,C_{1,2} \rangle$ generates no new group elements, and instead lowers the graph diameter by introducing additional edges between the set of vertices. Adding $P_1$ to the set $\langle H_1, \,H_2, \,C_{1,2} \rangle$ does generate additional elements---in fact $\langle H_1, \,H_2, \,P_1, \,C_{1,2} \rangle$ generates all of $\mathcal{C}_2$---but also lowers the Cayley graph diameter by adding additional edges.

We now discuss in depth how several subgroups are constructed, offering an explanation for order of each group seen in Table \ref{tab:OrbitLengthCliffordSubgroupNoRelations}. An extended version of Table \ref{tab:OrbitLengthCliffordSubgroupNoRelations}, containing the relations needed to present each subgroup, is given in Appendix \ref{ExtendedTableAppendix}. Additional Cayley graph illustrations are given in Appendix \ref{AdditionalGraphsAppendix}.

\paragraph{Single-Generator Subgroups:} The $\mathcal{C}_2$ subgroups generated by a single Clifford element, i.e. $\langle H_i \rangle, \,\langle P_i \rangle$, and $\langle C_{i,j} \rangle$, are completely described by Eqs. \eqref{HSquared}, \eqref{PFourth}, and \eqref{CSquared} respectively. Groups $\langle H_i \rangle$ and $\langle P_i \rangle$ were discussed in Section \ref{OneQubitSection}, and their Cayley graphs shown in Figure \ref{HadamardAndPhaseCayleyGraphs}. At two qubits we have the possibility of bi-local gates, such as $C_{i,j}$. Since $C_{i,j}^2 = \mathbb{1}$, as shown by Eq. \eqref{CSquared}, the group $\langle C_{i,j} \rangle$ is isomorphic to $\langle H_i \rangle$. 

\begin{figure}[h]
\begin{center}
\includegraphics[width=10.5cm]{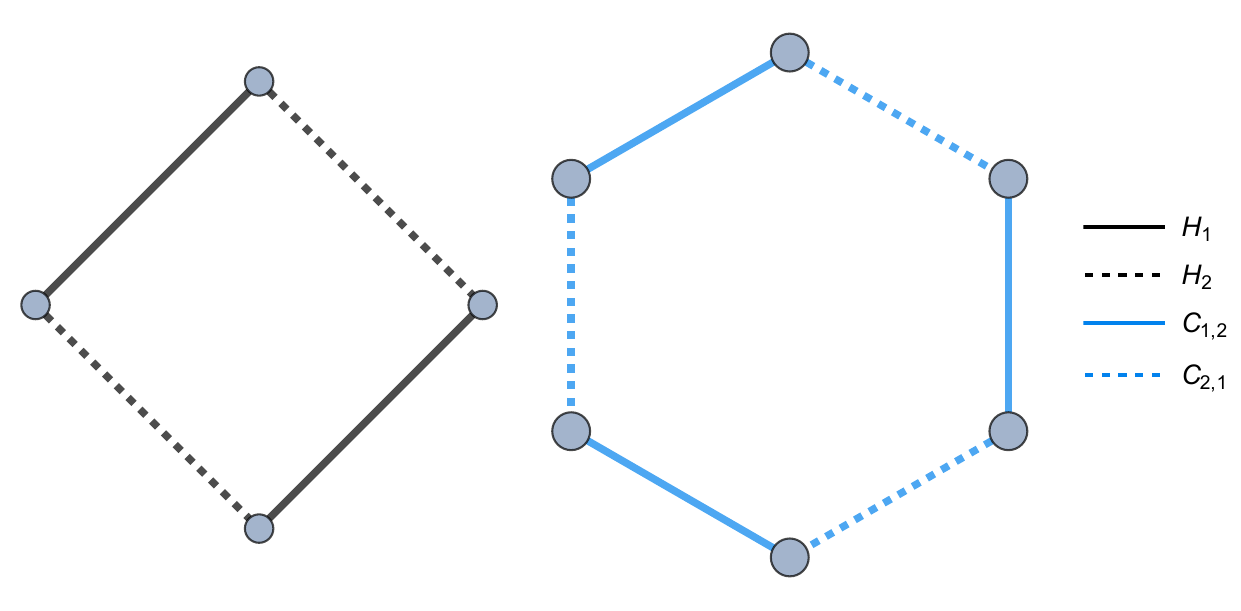}
\caption{Cayley graphs for subgroups $\langle H_1,H_2\rangle$ and $\langle C_{1,2},C_{2,1} \rangle$ arrange into square and hexagonal structures respectively. All edges in these figures are undirected since both $H$ and $C$ are their own inverse.}
\label{HBoxCHexPaper}
\end{center}
\end{figure}

\paragraph{Subgroups $\langle H_i,H_j \rangle$ and $\langle C_{i,j},C_{j,i} \rangle$:}
The subgroup generated by $\{H_i,H_j\}$ is completely described by Eqs. \eqref{HSquared} and \eqref{HhComm}. Since $H_i$ and $H_j$ commute for $i\neq j$, the group $\langle H_i,H_j \rangle$ has only $4$ elements, and its structure can be easily understood by examining the left image of Figure \ref{HBoxCHexPaper}. Similarly, the subgroup $\langle C_{i,j},C_{j,i} \rangle$ is described by Eqs. \eqref{CSquared} and \eqref{CcComm}. The elements $C_{i,j}$ and $C_{j,i}$ do not commute, but instead form the hexagonal structure to the right of Figure \ref{HBoxCHexPaper}.

\paragraph{Subgroup $\langle H_1, \,P_2, \,C_{2,1} \rangle$:} The subgroup generated by $\{H_1, \,P_2, \,C_{2,1}\}$ can be presented using Eqs. \eqref{HSquared}, \eqref{PFourth}, \eqref{HPComm}, \eqref{CSquared}, \eqref{HpComm}, \eqref{PCComm}, and \eqref{ChFourth}. Since $P_2$ commutes with both $H_1$ and $C_{2,1}$, all $P_2$ in a word can be pushed completely to one end, such that they occur either before or after all $H_1$ and $C_{2,1}$ operations. In this way, the elements of $\langle H_1, \,P_2, \,C_{2,1} \rangle$ can be constructed as products of some element from $\langle H_1, \,C_{2,1} \rangle$ with an element from the set $\{\mathbb{1},\,P_2,\,P_2^2,\,P_2^3\}$. Initially, this generates $16 \times 4 = 64$ words, however Eq. \eqref{ChFourth} demonstrates how $P_2^2$ can be built from $H_1$ and $C_{2,1}$. Therefore, all words containing $P_2^2$ or $P_2^3$ can be reduced to a shorter sequence, and the order $\langle H_1, \,P_2, \,C_{2,1} \rangle$ becomes $32$. 

\begin{figure}[h]
\begin{center}
\includegraphics[width=9cm]{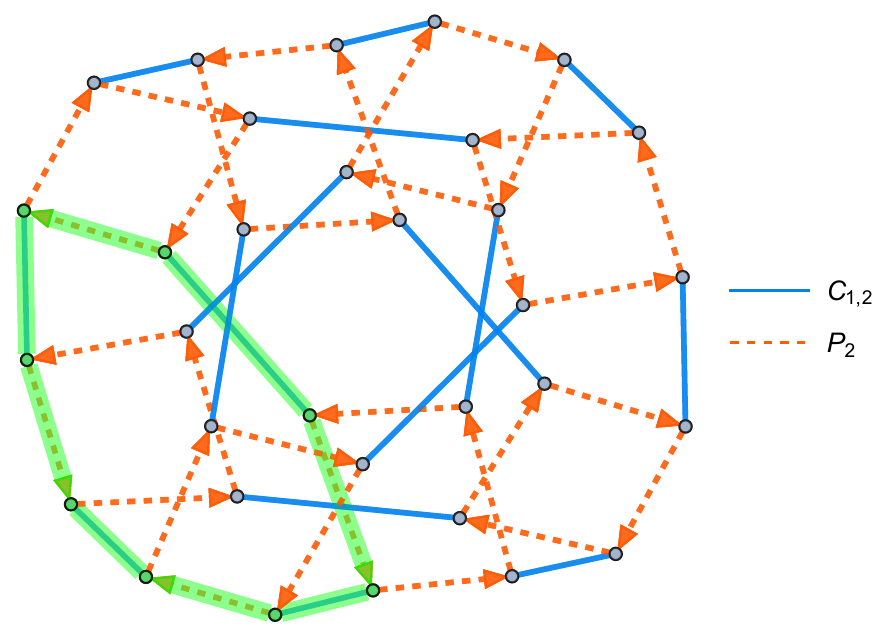}
\caption{Cayley graph of $\langle P_2,\,C_{1,2}\rangle$ subgroup, which is useful for visualizing relations such as $C_{i,j}P_jC_{i,j}P_j = P_jC_{i,j}P_jC_{i,j}$, highlighted in green.}
\label{C12P2CayleyGraph}
\end{center}
\end{figure}

\paragraph{Subgroup $\langle P_2,\,C_{1,2}\rangle$:} The subgroup generated by $\{ P_2,\,C_{1,2}\}$ can be built using Eqs. \eqref{PFourth}, \eqref{CSquared}, and \eqref{CpFourth}. Figure \ref{C12P2CayleyGraph} shows the Cayley graph for $\langle P_2,\,C_{1,2}\rangle$. We construct this subgroup by building words of alternating $C_{1,2}$ and $p$, where
\begin{equation}\label{PhaseGroup}
    p \in \{\mathbb{1},\,P_2,\,P_2^2,\,P_2^3\}.    
\end{equation}
 For clarity, we introduce the notation $\overline{p} \in \{P_2,\,P_2^2,\,P_2^3\}$. We generate all words containing up to $2$ CNOT operations, since the relation
 \begin{equation}\label{CpCpRelation}
     C_{i,j}P_jC_{i,j}P_j = P_jC_{i,j}P_jC_{i,j},
 \end{equation}
 derived in Eq. \eqref{CpCpRelationDerivation}, allows words with $3$ or more $C_{1,2}$ operations to be written as duplicates of words containing fewer $C_{1,2}$ operations.
    \begin{enumerate}
        \item For words containing $0$ $C_{1,2}$ operations, we have only the set $p$, containing $4$ unique elements.
        \item Words containing a $1$ $C_{1,2}$ operation have the form $pC_{1,2}p$, with full choice of $p$ on either side of $C_{1,2}$, giving $4 \times 4 = 16$ possible new elements.
        \item Words containing $2$ or more $C_{1,2}$ operations must alternate $C_{1,2}$ and $\overline{p}$ operations, or could otherwise be reduced by $(C_{1,2})^2 = \mathbb{1}$. Thus all $2$ $C_{1,2}$ words have the form $C_{1,2}\overline{p}C_{1,2}p$ (note that we never include $\mathbb{1}$ between $C_{1,2}$ operations as it could be carried through a $C_{1,2}$ to collapse the $C_{1,2}$ pair). We apply Eq. \eqref{CpCpRelation} to any word of the form $pC_{1,2}pC_{1,2}p$ to move all $p$ operations as far to the right as possible. In this way, we generate $3 \times 4 = 12$ new elements.
    \end{enumerate}
The above construction explicitly generates the $4+16+12 = 32$ elements of $\langle C_{1,2},\, P_2 \rangle$, each having one of the forms $\{p,\, pCp,\, C\overline{p}Cp\}$.

\paragraph{Subgroup $\langle H_1,\, P_2,\, C_{1,2} \rangle$:} The subgroup generated by $\{H_1,\, P_2, \,C_{1,2}\}$ can be built using Eqs. \eqref{HSquared}--\eqref{HPComm}, \eqref{CSquared}, \eqref{PpComm}, \eqref{HpComm}, \eqref{FourGenRelation}, \eqref{CNOTTransform}, \eqref{CcComm}, \eqref{PCComm}, and \eqref{ChFourth}. We will also use
\begin{equation}\label{CHpSquared}
    (C_{1,2}H_1P_2^2)^2 = (P_2^2H_1C_{1,2})^2,
\end{equation}
which can be derived from the relations \eqref{CSquared}--\eqref{ChFourth}. 

As with $\langle P_2, \,C_{1,2} \rangle$, we construct $\langle H_1,\,P_2, \,C_{1,2} \rangle$ by building words of alternating $H_1$ and one element from $\{P_2, \,C_{1,2}\}$, since $(H_1)^2 = \mathbb{1}$. We only need to construct all words containing up to $5$ $H_1$ operations, since words with $6$ $H_1$ operations can be reduced using $(H_1C_{1,2})^8 = \mathbb{1}$ or $(H_1C_{1,2}P_2)^6 = (P_2H_1C_{1,2})^6 = \omega^6$. The full construction of $\langle H_1,\,P_2, \,C_{1,2} \rangle$ is given in Appendix \ref{SubgroupPresentationAppendix}. 

Appending $H_2$ to the generating set $\{H_1, \,P_2, \,C_{1,2}\}$ results in a factor of $10$ more elements, giving the full group $\mathcal{C}_2$. Adding more generators to $\{H_1, \,H_2, \,P_2, \,C_{1,2}\}$ does not add more group elements, and instead lowers the graph diameter. In fact, the set $\{H_1, \,H_2, \,P_2, \,C_{1,2}\}$ is a minimal generating set%
\footnote{There exist generating sets for $\mathcal{C}_2$ with fewer elements which involve composite Clifford operations e.g. the set $\{H_1P_1, \,H_2P_2, \,C_{1,2}\}$.} %
for $\mathcal{C}_2$ with the generators $\{H_1, \,H_2, \,P_1, \,P_2, \,C_{1,2}, \,C_{2,1}\}$. 

\begin{figure}[h]
\begin{center}
\includegraphics[width=10cm]{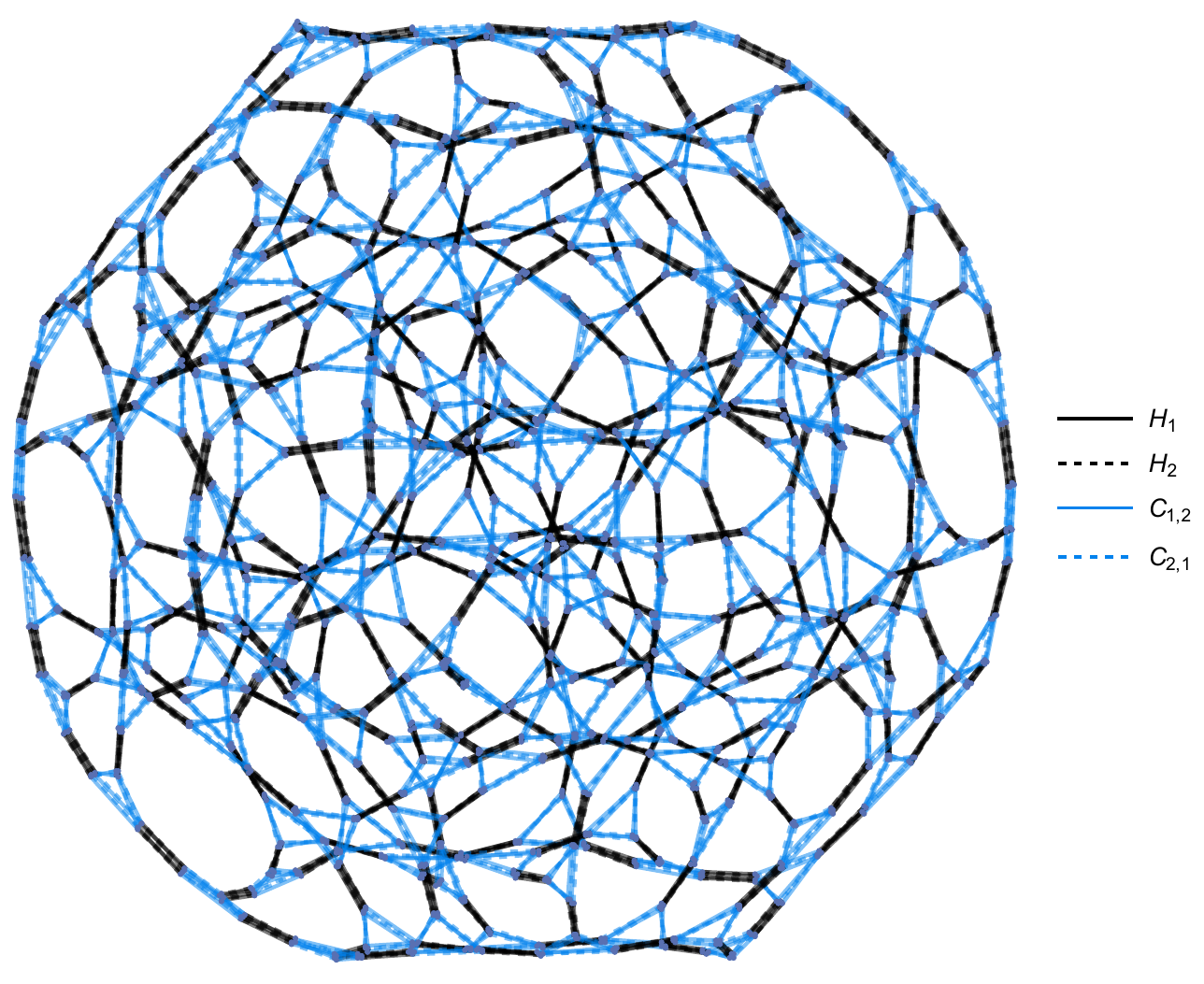}
\caption{The Cayley graph for $\mathcal{C}_2$ subgroup $\langle H_1,\,H_2,\,C_{1,2},\,C_{2,1}\rangle$. This graph has $2304$ vertices, and displays the orbit of an arbitrary quantum state under the action of $\langle H_1,\,H_2,\,C_{1,2},\,C_{2,1}\rangle$.}
\label{HhCcCayleyGraph}
\end{center}
\end{figure}

\paragraph{Subgroup $\langle H_1, \,H_2, \,C_{1,2}, \,C_{2,1} \rangle$:} The group generated by $\{H_1, \,H_2, \,C_{1,2},\,C_{2,1} \}$ can be understood from Eqs. \eqref{HSquared}, \eqref{PFourth}, \eqref{HPComm}, \eqref{CSquared}, \eqref{CcComm}, and \eqref{ChFourth}. We additionally make use of the identity,
\begin{equation}\label{HadamardTransform}
    C_{j,i}C_{i,j}C_{j,i}H_iC_{j,i}C_{i,j}C_{j,i} = H_j,
\end{equation}
which transforms a Hadamard using a sequence of CNOT operations. Initially, we might assume this group is the direct product%
\footnote{While $\langle H_1, \,H_2, \,C_{1,2}, \,C_{2,1} \rangle$ is not a direct product, one example which is a direct product is $\langle H_1, \,H_2, \,P_1,\,P_2 \rangle  = \langle H_1, \,P_1 \rangle \times \langle H_2,\,P_2 \rangle $, which has $192^2/8=4608$ elements.} %
of $\langle H_1, \,C_{1,2} \rangle$ and $\langle H_2, \,C_{2,1} \rangle$, thereby having $256$ elements. However, Eq. \eqref{ChFourth} importantly demonstrates how a sequence of $H$ and $C$ operations can create $P_i^2$. This generates a factor of $9$ more elements, for a total of $2304$. Furthermore, since Eq. \eqref{CNOTTransform} offers a way to construct $C_{2,1}$ as the product of $H_1, \,H_2,$ and $C_{1,2}$, the subgroup $\langle H_1, \,H_2, \,C_{1,2}, \,C_{2,1}\rangle$ can be minimally generated from sets $\{H_i,\,H_j,\,C_{i,j}\}$ or $\{H_i,\,C_{i,j},\,C_{j,i}\}$. Figure \ref{HhCcCayleyGraph} shows the Cayley graph for $\langle H_1, \,H_2, \,C_{1,2}, \,C_{2,1} \rangle$.

Through building this presentation and constructing subgroups in detail, we have developed a functional understanding of $\mathcal{C}_2$. We identified a collection of Clifford group relations which are independent of the state set being acted on. This state-independent description will allow us to extend an analysis beyond the set of stabilizer states, and to explore action of the Clifford group on arbitrary quantum states. By systematically constructing all words in a subgroup, we were able to highlight exactly how our relations transform Clifford strings. We found additional relations, such as $C_{i,j}P_jC_{i,j}P_j = P_jC_{i,j}P_jC_{i,j}$ and $C_{j,i}C_{i,j}C_{j,i}H_iC_{j,i}C_{i,j}C_{j,i} = H_j$, which are not included in our presentation, but can be derived from relations \eqref{CSquared}--\eqref{ChFourth}. These auxiliary relations proved useful for understanding why certain sequences of Clifford gates are non-trivially equivalent to others, as well as how entanglement entropy evolves through Clifford circuits.

Constructing the Cayley graph for each subgroup further illustrated the structure of $\mathcal{C}_2$ and its subgroups. These graphs enabled us to visualize the operator relations that were used to build each subgroup. Computing the Cayley graph diameter, and observing its change after adding generators or quotienting by $\omega$, offered additional intuition for $\mathcal{C}_2$ subgroup connectivity. These Cayley graphs will constitute a state-independent starting point for constructing reachability graphs in the following section. By considering quotient spaces of this purely group-theoretic structure, we are able to analyze the orbit of arbitrary quantum states under a selected gate set. Furthermore, by understanding how this quotient protocol modifies a Cayley graph we are able to build alternative graphs that can track and bound the evolution of certain system properties, such as entanglement entropy \cite{Keeler:2023shl}.

\section{Reachability Graphs as Cayley Graph Quotients}\label{ReachabilityGraphsFromCayleyGraphs}

We now generalize the notion of reachability graphs by constructing them as quotient spaces of Cayley graphs. We define equivalence classes on group elements by their congruent action on a chosen state. We demonstrate how identifying vertices in a Cayley graph collapses its structure to a state reachability graph. Starting with the state-independent Cayley graph, we are able to strictly bound the orbits for different states under select sets of gates.

In each example below, we first quotient Cayley graphs by global phase, then by the stabilizer subgroup of a chosen state. We explicitly compute quotients of $\mathcal{C}_1$ and $\mathcal{C}_2$ Cayley graphs which yield familiar reachability graphs for one and two qubit stabilizer states. Restricting to the subgroup $\langle H_1,\,H_2,\,C_{1,2},\,C_{2,1} \rangle < \mathcal{C}_2$, which we denote $\HC$ going forward, we recover the reachability graphs studied in \cite{Keeler2022}.

By adding $P_1$ and $P_2$ to the set $\{ H_1,\,H_2,\,C_{1,2},\,C_{2,1} \}$, we consider the full action of $\mathcal{C}_2$. We observe how the addition of these two phase gates ties disconnected $\HC$ subgraphs together. Finally we apply our generalized understanding of $\mathcal{C}_2$ operators to extend beyond the set of stabilizer states, and generate $\HC$ orbits for non-stabilizer states.

\subsection{Quotient by Global Phase}\label{QuotientByPhase}

In this section, we define a procedure to quotient%
\footnote{Formally we are building the map $\mathcal{Q}: G \rightarrow G/N$, which takes elements of a group $G$ into a set of equivalence classes $G/N$. The set of equivalence classes is fixed by choice of congruence relation, e.g. congruence up to action by $\omega^n$. When the group $N$ is normal in $G$, the set $G/N$ is a formal quotient group of $G$.} %
by elements which act as a phase on the group. When building reachability graphs from Cayley graphs we always quotient first by the group element $\omega = (H_1P_1)^3$, as in Eq. \eqref{HPComm}, since quantum states can only be operationally distinguished up to global phase. Accordingly, all graph quotients we construct going forward are quotients by the product $\langle \omega\rangle \times H$.

We begin by explicitly building the quotient of $\mathcal{C}_1/\langle \omega \rangle$. As discussed in Section \ref{OneQubitSection}, the group $\mathcal{C}_1$ is generated by $\{H_1,\,P_1\}$ and contains $192$ elements. When quotienting by $\omega$, we identify together all elements of $\mathcal{C}_1$ that are equivalent up to powers of $\omega$. For $g_1,\,g_2 \in \mathcal{C}_1$,
\begin{equation}
    g_1 \equiv g_2 \textnormal{ if } g_1 = \omega^{n\Mod{8}}g_2.
\end{equation}
This identification defines the normal subgroup $\langle \omega \rangle \lhd \mathcal{C}_1$, where
\begin{equation}
    \langle \omega \rangle \equiv \{\mathbb{1},\,\omega,\,\omega^2,\,\omega^3,\,\omega^4,\,\omega^5,\,\omega^6,\,\omega^7\},
\end{equation}
and allows us to construct the quotient group $\bar{\mathcal{C}_1} \equiv \mathcal{C}_1/\langle \omega \rangle$. 

The quotient $\bar{\mathcal{C}_1}$ consists of $24$ equivalence classes of $8$ elements each. This is a factor of $8$ reduction in group order, from $192$ to $24$, as shown in the $\{H_1, \,P_1\}$ row of Table \ref{tab:OrbitLengthCliffordSubgroupNoRelations}. All elements of each class are equivalent up to powers of $\omega$. The $24$ equivalence classes can be represented by elements of the form
\begin{equation}\label{NonTrivEqClasses}
    \begin{split}
        \{p,\, pH_1p,\, H_1P_1^2H_1p\},
    \end{split}
\end{equation}
where $p \in \{\mathbb{1},\, P_1, \, P_1^2,\, P_1^3\}$ as defined in Eq. \eqref{PhaseGroup}.

\begin{figure}[h]
        \centering        
        \includegraphics[width=13cm]{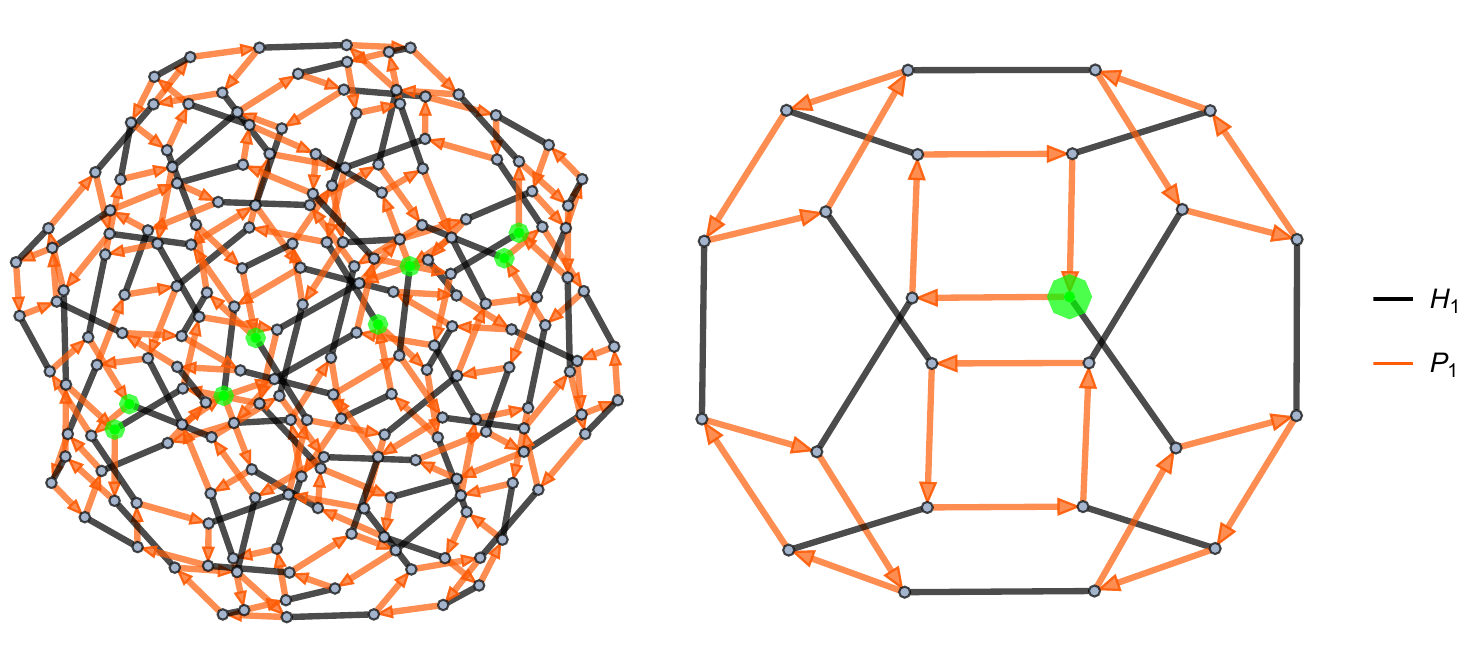}
        \caption{Cayley graph of $\mathcal{C}_1$  before and after quotienting by $\omega$. The $192$ vertices in the $\mathcal{C}_1$ Cayley graph collapse to $24$ vertices in the $\bar{\mathcal{C}_1}$ quotient graph. Every edge in the quotient graph represents the $8$ edges $\omega^nH_1$ and $\omega^nP_1$. One set of $8$ vertices, which are identified to a single vertex under this quotient, is highlighted in green.}
        \label{C1Quotient}
    \end{figure}

Quotienting $\mathcal{C}_1$ by $\langle \omega \rangle$ likewise modifies the $\mathcal{C}_1$ Cayley graph by gluing together all vertices that represent operators in the same equivalence class. Figure \ref{C1Quotient} shows the Cayley graph of $\mathcal{C}_1$ before and after modding out by $\omega$. Each vertex in the $\bar{\mathcal{C}_1}$ graph represents $8$ elements of $\mathcal{C}_1$, collapsing the $192$ vertices of the $\mathcal{C}_1$ Cayley graph down to $24$. Every $H_1$ edge in the contracted graph represents the $8$ operators $\omega^nH_1$, and similarly for $P_1$.

We have described a procedure for quotienting by $\omega$, which acts as a global phase. In the following sections, we will first quotient by $\omega$ when constructing reachability graphs. Quotienting a group by $\omega$ contracts the Cayley graph, by identifying vertices which represent elements equivalent up to $\omega$. As we highlight below, a similar graph contraction will yield reachability graphs as Cayley graph quotients, which we will highlight further in the following sections.

\subsection{Quotient by Stabilizer Subgroup}

In this section we show how to generate cosets by the stabilizer subgroup of a chosen state, and how to construct the state's reachability graph as a quotient space of the group Cayley graph. While Cayley graphs offer a state-independent description of a group, the orbit of a particular state under that group action is state-dependent. Our quotient procedure defines the collapse of a group Cayley graph into a subgraph which gives the reachability graph for a chosen state.

A state's reachability graph displays the evolution of that state under some chosen set of quantum gates. Since we are deriving each reachability graph from the Cayley graph of a group, this procedure can be applied to any\footnote{We do require that the state should be thought of as a state on $n$ qubits, i.e. with a fixed factorization in a fixed $2^n$-dimensional Hilbert space.} chosen quantum state on which the group acts.

The general procedure is to identify a group $G$ which acts on a Hilbert space $\Hil$, as well as a choice of generators for $G$. We first quotient $G$ by the global phase%
\footnote{For the general $G$, $\omega$ can be any element that acts as a root of unity times the identity operator.  For the Clifford subgroups we study here, $\omega$ will be an eighth root of unity.}
$\omega$, giving the quotient group $\bar{G} = G/\langle \omega \rangle$. Each element of $\bar{G}$ is isomorphic to the equivalence class $\omega^ng \in G$, for some $g \in G$. We then identify a state $\ket{\psi} \in \Hil$, which selects the stabilizer subgroup $\Stab_{\bar{G}}(\ket{\psi})$ that we will use to build left cosets of $\bar{G}$. Since $\Stab_{\bar{G}}(\ket{\psi})$ is a subgroup, we generate the left coset space $\bar{G}/\Stab_{\bar{G}}(\ket{\psi})$ by computing all
\begin{equation}
    g\cdot\Stab_{\bar{G}}(\ket{\psi}) \quad \forall g \in \bar{G}.
\end{equation}

Constructing the above coset space generates a set of equivalence classes on $G$, with elements of each class congruent in their action on $\ket{\psi}$. To map elements between different equivalence classes we define the function $f: \Stab_{\bar{G}}(\ket{\psi}) \rightarrow \Stab_{\bar{G}}(\ket{\phi})$, where
\begin{equation}
    f(g) = hg^{-1}, \quad \forall \,g \in G_{\ket{\psi}},\, h \in G_{\ket{\phi}}.
\end{equation}
For example, to transform $P_1 \in \Stab_{\bar{\mathcal{C}_2}}(\ket{00})$ to $H_1P_1^2H_2P_2^2 \in \Stab_{\bar{\mathcal{C}_2}}(\ket{GHZ}_2)$ we apply the sequence $H_1P_1^2H_2P_2^2P_1^{-1}$. 

As an illustration of this procedure, we construct the left cosets of $\mathcal{C}_1$ by $\langle \omega \rangle \times \Stab_{\mathcal{C}_1}(\ket{0})$. We first build the quotient group $\bar{\mathcal{C}}_1 = \mathcal{C}_1/\langle \omega \rangle$ as detailed in Section \ref{QuotientByPhase}. We then identify the stabilizer group $\Stab_{\mathcal{C}_1}(\ket{0}) < \bar{\mathcal{C}}_1$, which comprises the $4$ elements that stabilize $\ket{0}$, i.e.
\begin{equation}
    \Stab_{\mathcal{C}_1}(\ket{0}) = \{\mathbb{1}, \,P_1, \,P_1^2, \,P_1^3\}.
\end{equation}
Build all left cosets of $\bar{\mathcal{C}}_1$ by $\Stab_{\mathcal{C}_1}(\ket{0})$ then gives a set of $6$ equivalence classes, with a representative element from each class being
\begin{equation}\label{RepresentativesC1}
    \{\mathbb{1},\, pH_1,\, H_1P_1^2H_1\},
\end{equation}
with $p \in \{\mathbb{1},\, P_1, \, P_1^2,\, P_1^3\}$ as before. The elements of each equivalence class are identified by multiplying each representative in Eq. \eqref{RepresentativesC1} by the $4$ elements of $\Stab_{\mathcal{C}_1}(\ket{0})$. 

\begin{figure}[h]
    \centering
		\begin{overpic}[width=12.5cm]{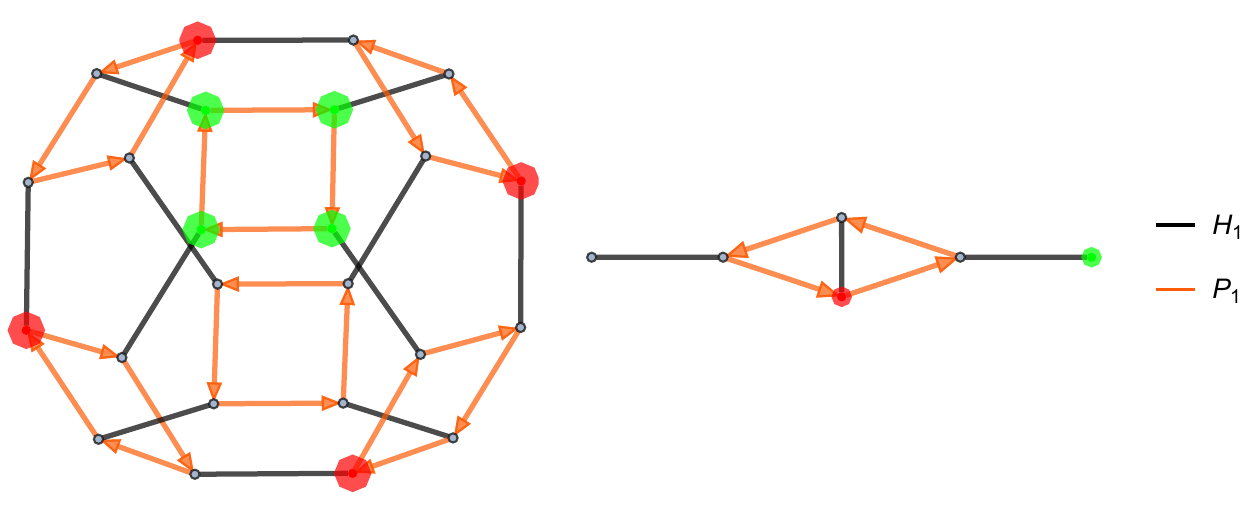}
		\put (79.5,22){\footnotesize{$\ket{0} \searrow$}}
		\put (67.5,14.5) {\footnotesize{$\nwarrow \ket{-i}$}}
        \end{overpic}
        \caption{Left cosets of $\mathcal{C}_1$ by $\Stab_{\mathcal{C}_1}(\ket{0})$, the stabilizer subgroup of $\ket{0}$. The $\mathcal{C}_1$ Cayley graph on the left collapses to a $6$-vertex reachability graph on the right. Four green vertices identify to a single vertex representing the equivalence class of $\Stab_{\mathcal{C}_1}(\ket{0})$, while four red vertices likewise identify to a vertex for $\Stab_{\mathcal{C}_1}(\ket{-i})$.}
		\label{VacuumQuotient}
	\end{figure}

We build the graph corresponding to $\bar{\mathcal{C}}_1/\Stab_{\mathcal{C}_1}(\ket{0})$ by assigning a vertex to each of the $6$ equivalence classes. Figure \ref{VacuumQuotient} shows the $\mathcal{C}_1$ Cayley graph before and after modding by $\langle \omega \rangle \times \Stab_{\mathcal{C}_1}(\ket{0})$. The $192$-vertex $\mathcal{C}_1$ Cayley graph is reduced to a graph with $6$ vertices, which is isomorphic to the complete reachability graph for single-qubit stabilizer states.

We have demonstrated a procedure for building the left coset space of Clifford groups and subgroups for the stabilizer subgroup of a quantum state. We illustrated how Cayley graphs quotient to state reachability graphs under this equivalence relation. We will now use this protocol to explore subgroups of $\mathcal{C}_2$.

\subsection{Stabilizer Restricted Graphs from $\langle H_i,\,H_j,\,C_{i,j},\,C_{j,i} \rangle$ Quotients}\label{HhCcStabilizerQuotients}

In \cite{Keeler2022} we constructed and analyzed reachability graphs under the action of $\mathcal{C}_2$ subgroups. We termed these restricted graphs, and focused on the subgroup\\ $\HC \equiv \langle H_1,\,H_2,\,C_{1,2},\,C_{2,1}\rangle$. Since entanglement entropy among stabilizer states is modified by, at most, bi-local action, this subgroup gives useful insight into stabilizer entanglement. We now generalize the construction of restricted graphs in \cite{Keeler2022} by constructing the reachability graphs as quotient spaces of Cayley graphs. We specifically reproduce all $\HC$ restricted graphs that arise for stabilizer states, then use our model to explore the orbit of non-stabilizer states as well. 

The quotiented Cayley graphs we construct, in addition to representing a particular left coset space, are isomorphic to state reachability graphs. As defined in \cite{Keeler2022}, the vertices of reachability graphs represent states in a Hilbert space, while edges represent gates acting which transform these states. Vertices in the quotient space of a Cayley graph represent equivalence classes of group elements, defined by their orbit with respect to a chosen subgroup, while edges represent sets of generators. Going forward, we refer to Cayley graph quotients as reachability graphs and note the distinction when necessary.

All stabilizer states can reached by acting on $\ket{0}^{\otimes n}$ with $\mathcal{C}_n$. Acting on $\ket{0}^{\otimes n}$ with the subgroup $\HC$ generates $24$ stabilizer states, including all measurement states of the computational basis. Every state in the orbit of $\ket{0}^{\otimes n}$ is stabilized by $48$ elements of $\HC$.

\begin{figure}[h]
    \centering
        \begin{overpic}[width=12.5cm]{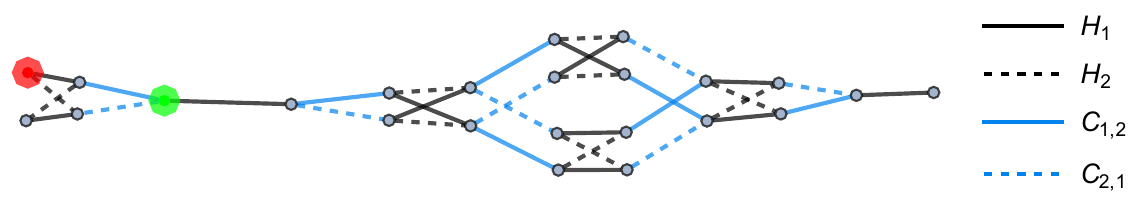}
		  \put (3.25,14) {\footnotesize{$\swarrow \ket{00}$}}
		  \put (15.2,6) {\footnotesize{$\nwarrow \ket{GHZ}_2$}}
        \end{overpic}
    \caption{Quotient space of $\HC$ Cayley graph after modding out by the stabilizer subgroup of $\ket{00}$. The equivalence class of $\Stab_{\HC}(\ket{00})$ is highlighted in red, while the equivalence class of $\Stab_{\HC}(\ket{GHZ}_2)$ is highlighted in green.}
    \label{g24CayleyQuotient}
    \end{figure}

Figure \ref{g24CayleyQuotient} shows the $\HC$ Cayley graph after quotienting by the stabilizer subgroup for $\ket{0}^{\otimes n}$, specifically for the $2$-qubit example $\ket{00}$. Since the stabilizer subgroup is preserved when tensoring on additional qubits to the system, this $24$-vertex graph likewise displays the orbit for any product of $\ket{00}$ with any $(n-2)$-qubit state. Each vertex in Figure \ref{g24CayleyQuotient} represents the stabilizer subgroup for one state in the orbit of $\ket{00}$ under $\HC$.

In general any product state, as well as states with no entanglement between the first two qubits and the other $n-2$, will have either the $24$-vertex reachability graph in Figure \ref{g24CayleyQuotient} or the $36$-vertex reachability graph, also defined in \cite{Keeler2022}, that appears below in Figure \ref{HhCcPhaseOverlay2}. 

\begin{figure}[h]
    \centering
		\begin{overpic}[width=11cm]{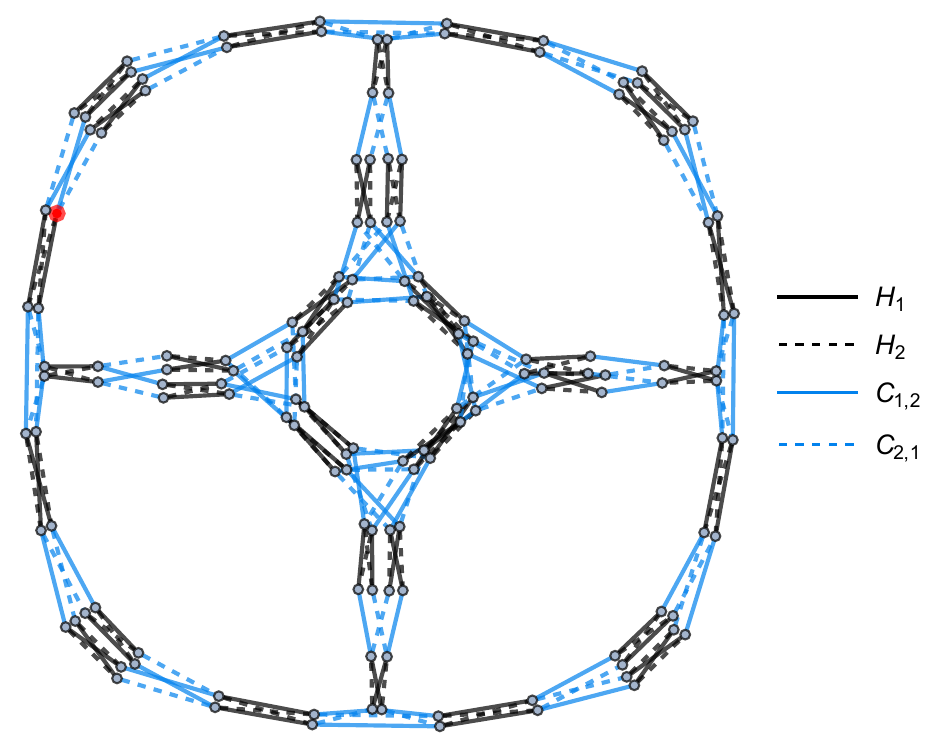}
		\put (7.5,56.6) {\footnotesize{$\leftarrow \ket{GHZ}_3$}}
        \end{overpic}
        \caption{Quotient of $\HC$ Cayley graph after modding by stabilizer subgroup for $\ket{GHZ}_3$. Red vertex gives the equivalence class of $\HC$ elements that stabilize $\ket{GHZ}_3$.}
    \label{g144CayleyQuotient}
	\end{figure}

For additional entangled states which arise at higher qubit number, new reachability graph structures appear when acting with $\HC$. Figure \ref{g144CayleyQuotient} shows a quotient space of the $\HC$ Cayley graph after modding out by the stabilizer subgroup for $\ket{GHZ}_3 \equiv \ket{000} + \ket{111}$. The orbit of $\ket{GHZ}_3$ under $\HC$ reaches $144$ states, each of which is stabilized by $8$ elements of $\HC$.

\begin{figure}[h]
    \centering
		\begin{overpic}[width=11cm]{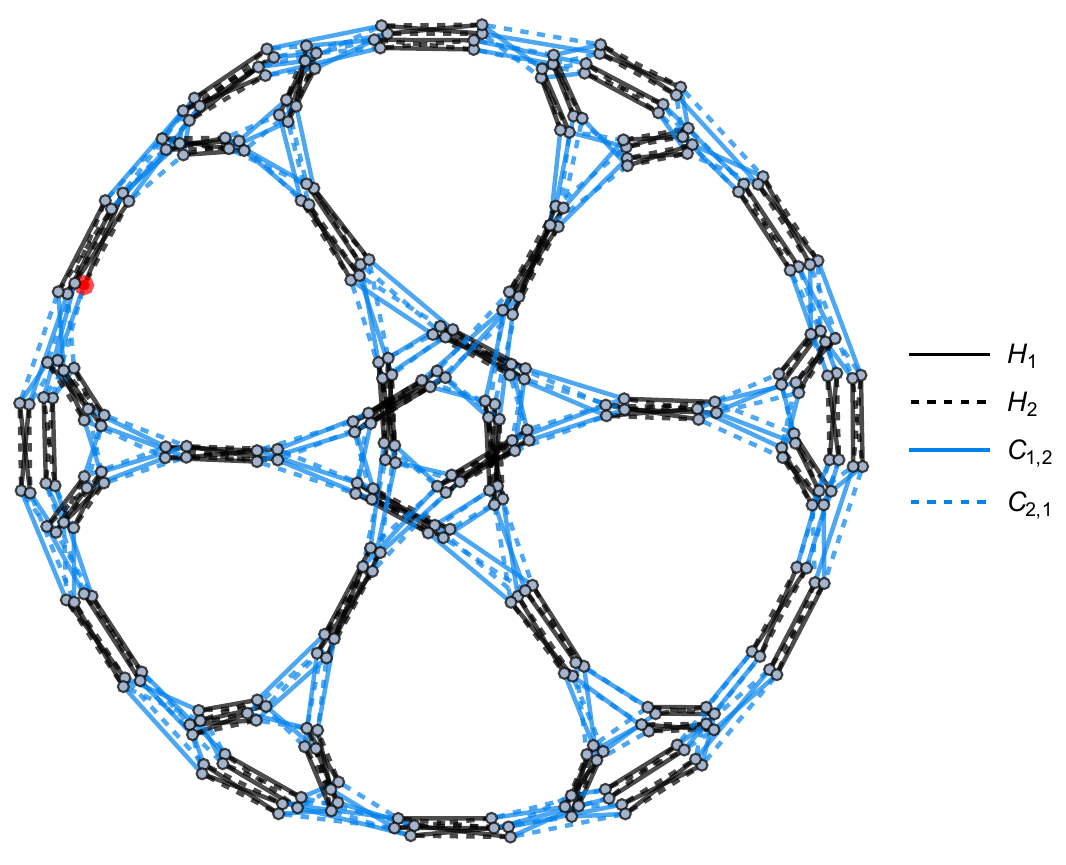}
		\put (9,53.5) {\footnotesize{$\leftarrow C_{3,2}\ket{i1+}$}}
        \end{overpic}
        \caption{Orbit of stabilizer states which are stabilized by $4$ elements of $\HC$. This graph has different topology from the $288$-vertex graph in Figure \ref{D31HhCcGraph}. Elements of $\HC$ that stabilize $C_{3,2}\ket{i1+} \equiv \ket{010} + i\ket{011} + \ket{100} + i\ket{101}$ are represented by the red vertex.}
    \label{g288CayleyQuotient}
	\end{figure}

Also at three qubits, there exist stabilizer states which are stabilized by only $4$ elements of $\HC$. In Section $4$ of \cite{Keeler2022}, we defined a lifting procedure which allows us to find an example state in each stabilizer reachability graph. To identify a state that is stabilized by $4$ elements of $\HC$, we act with $C_{3,2}$ on the product state $\ket{i} \otimes \ket{1} \otimes \ket{+}$. The resultant state $\ket{010} + i\ket{011} + \ket{100} + i\ket{101}$ is stabilized by the elements
\begin{equation}\label{g288StabGroup}
    \{\mathbb{1},\, H_2(C_{1,2}H_1)^4,\, (C_{1,2}H_1)^4H_2,\, \left((C_{1,2}H_1)^3C_{1,2}H_2\right)^2\}.
\end{equation}
Figure \ref{g288CayleyQuotient} shows the quotient space of $\HC$ after modding out by the stabilizer subgroup for $C_{3,2}\ket{i1+} \equiv \ket{010} + i\ket{011} + \ket{100} + i\ket{101}$.

\begin{figure}[h]
        \centering
        \includegraphics[width=11cm]{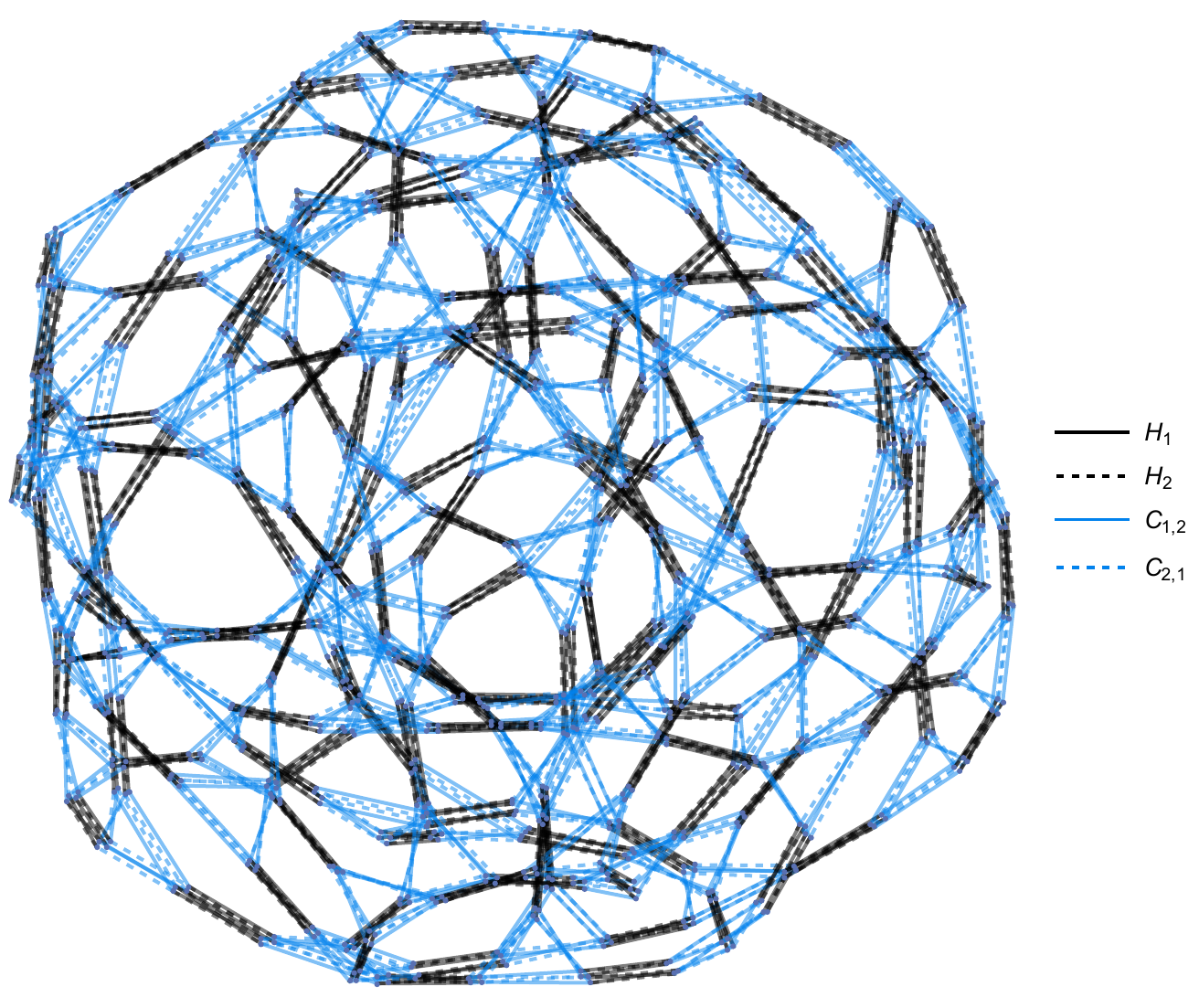}
        \caption{Orbit for states stabilized by only $\mathbb{1}$ in $\HC$. This $1152$-vertex graph gives the orbit of a generic quantum state under action of $\HC$.}
    \label{g1152CayleyQuotient}
    \end{figure}

Finally, there are stabilizer states which are only stabilized by $\mathbb{1}$ in $\HC$. Figure \ref{g1152CayleyQuotient} illustrates the $1152$-vertex reachability graph for such states. This graph represents the largest possible orbit of any quantum state under $\HC$, since all states are trivially stabilized by $\mathbb{1}$. Figure \ref{g1152CayleyQuotient} is first observed at four qubits. 

By taking quotients of the $\HC$ Cayley graph, we have reproduced all stabilizer reachability graphs found in \cite{Keeler2022} under the action of this subgroup. We demonstrated that the largest such subgraph contains $1152$ vertices, by representing the orbit of states which are stabilized by only the identity in $\HC$, in agreement with \cite{Keeler2022}. In the following subsection we add $P_1$ and $P_2$ back into our generating set, and study the action of the full group $\mathcal{C}_2$. We will show how the addition of these two phase gates does not generate any additional graphs, but instead connects the existent structures above. We also discover new reachability graphs that arise from quotienting $\HC$ Cayley graphs by stabilizer subgroups of non-stabilizer states.

\subsection{Full $\mathcal{C}_2$ Action}

Adding $P_1$ and $P_2$ to the set $\{H_1,\, H_2,\, C_{1,2},\, C_{2,1}\}$ generates the full group $\mathcal{C}_2$, which contains $92160$ elements. The Cayley graph for $\mathcal{C}_2$ after quotienting by $\langle \omega \rangle$ will accordingly have $11520$ vertices, as seen in the last four lines of Table \ref{tab:OrbitLengthCliffordSubgroupNoRelations}. Similarly the reachability graph for any $n$-qubit state stabilized by only $\mathbb{1}$ in $\mathcal{C}_2$ will have $11520$ vertices. By first considering the action of $\langle H_1,\, H_2,\, C_{1,2},\, C_{2,1}\rangle$ on a set of states, followed by the action of $P_1$ and $P_2$, we observe how the reachability graphs from Section \ref{HhCcStabilizerQuotients} are connected. 

\begin{figure}[h]
        \centering
        \begin{overpic}[width=11cm]{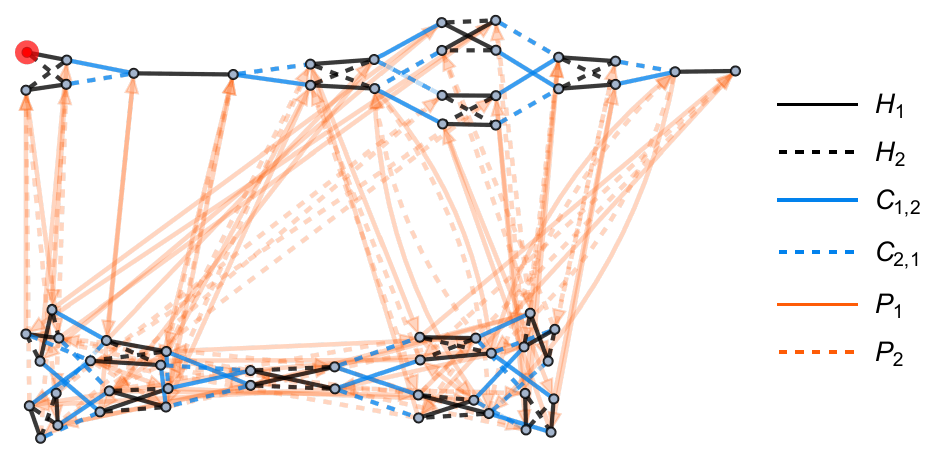}
            \put (4,45.5) {\footnotesize{$\swarrow \ket{00}^{\otimes n}$}}
        \end{overpic}
        \caption{Acting with $P_1$ and $P_2$ on all states in the $\HC$ orbit of $\ket{00}\otimes \ket{\psi}_{n-2}$ connects the $24$-vertex reachability graph from Figure \ref{g24CayleyQuotient} to a graph of $36$ vertices. Together these two graphs show the $\mathcal{C}_2$ orbit of any product state, and all states with no entanglement between the first two qubits and the remaining $n-2$ qubits.}
    \label{HhCcPhaseOverlay2}
    \end{figure}

\begin{figure}[h]
        \centering
        \includegraphics[width=12.5cm]{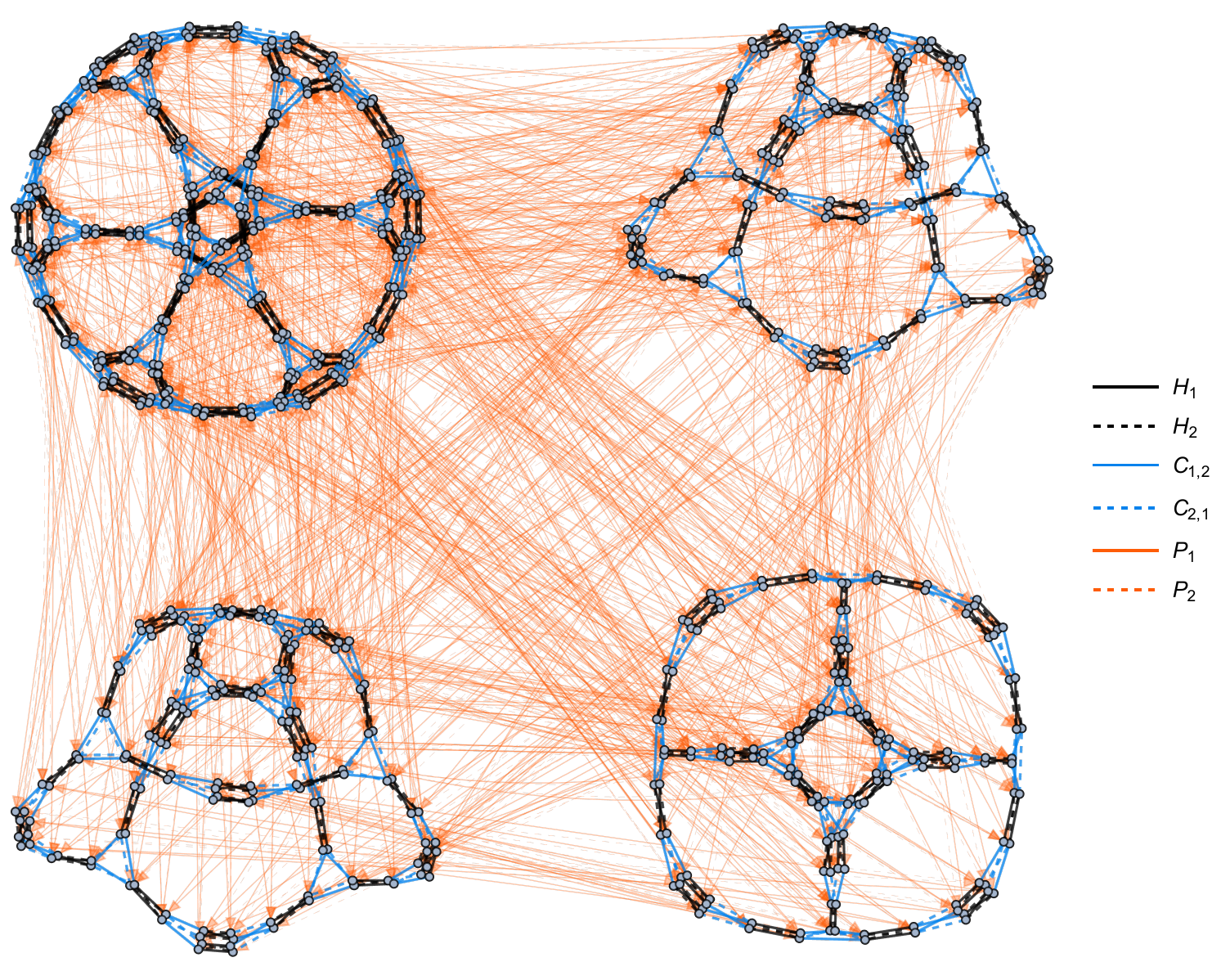}
        \caption{Three copies of the $144$-vertex reachability graph in Figure \ref{g144CayleyQuotient} connect to a single copy of the $288$-vertex graph in Figure \ref{g288CayleyQuotient}, after acting with $P_1$ and $P_2$.}
    \label{HhCcPhaseOverlay}
    \end{figure}

To illustrate how phase gates tie $\langle H_1,\, H_2,\, C_{1,2},\, C_{2,1}\rangle$ reachability graphs together, we first consider the orbit of $\ket{0}^{\otimes n}$ shown in Figure \ref{g24CayleyQuotient}. Acting with $P_1$ and $P_2$ on all states in this orbit connects the $24$-vertex reachability graph to a $36$-vertex graph, as in Figure \ref{HhCcPhaseOverlay2}. These two graphs combine to give the orbit of any pure state under the action of $\mathcal{C}_2$, as well as any state%
\footnote{Figure \ref{HhCcPhaseOverlay2} actually shows the orbit of any $n$-qubit state with no entanglement between one pair of qubits and the remaining $n-2$ qubits, since qubits $1$ and $2$ can be exchanged, without loss of generality, with any qubits in both the state and the Clifford subgroup.}
with only entanglement among its first two qubits. In both Figure \ref{HhCcPhaseOverlay2} and Figure \ref{HhCcPhaseOverlay} we have removed all ``trivial loops'', that is all edges which map a vertex back to itself, as these loops represent a stabilizing action on the vertex.

Similarly at higher qubit number, phase gates on the first two qubits tie together the larger $\langle H_1,\, H_2,\, C_{1,2},\, C_{2,1}\rangle$ reachability graphs. Acting with $P_1$ and $P_2$ on states in the stabilizer $288$-vertex graph, seen in Figure \ref{g288CayleyQuotient}, will sometimes act trivially, sometimes map the state to another in the $288$-vertex graph, and sometimes map it to one of three $144$-vertex graphs. Figure \ref{HhCcPhaseOverlay} depicts how these four graphs are connected via phase operations, where again trivial loops have been removed. 

\begin{figure}[h]
    \centering
 \includegraphics[width=8cm]{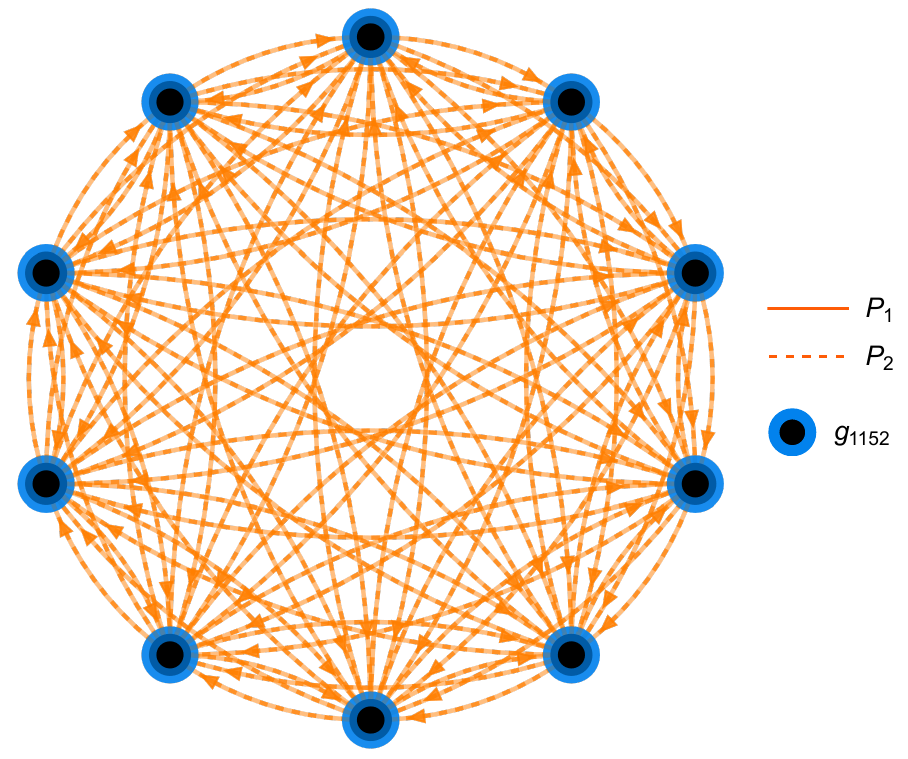}
    \caption{Acting $P_1$ and $P_2$ on states in $g_{1152}$ attaches the graph to $9$ isomorphic copies of itself, forming a completely-connected $10$-vertex graph. Each vertex represents one copy of $g_{1152}$ (Figure \ref{g1152CayleyQuotient}) and every edge is a set of phase gates which connect $g_{1152}$ graphs.}
    \label{Connected1152}
\end{figure}

The largest reachability graph under the action of $\HC$ contains $1152$ vertices, and is depicted in Figure \ref{g1152CayleyQuotient}. This reachability graph, which we term $g_{1152}$, gives the orbit of states which are stabilized by only the identity in $\HC$. Acting with $P_1$ and $P_2$ on every state in $g_{1152}$ either connects the $1152$-vertex graph to itself, or maps to one of its $9$ isomorphic copies. Figure \ref{Connected1152} shows how these $10$ copies of $g_{1152}$ are symmetrically attached via phase operations. Upon acting with $P_1$ and $P_2$ the resulting structure forms a completely-connected graph of $10$ vertices, where each vertex actually represents a $g_{1152}$ graph. 

We have examined the full action of $\mathcal{C}_2$ by acting with $P_1$ and $P_2$ on states in $\HC$ orbits. We demonstrated how the addition of these two phase gates ties together the $\HC$ reachability graphs shown in Section \ref{HhCcStabilizerQuotients}. Specifically we observed how the $24$-vertex reachability graph in Figure \ref{g24CayleyQuotient} connects to another graph of $36$ vertices. Meanwhile, three copies of the $144$-vertex graph in Figure \ref{g144CayleyQuotient} connect to a single copy of the $288$-vertex graph in Figure \ref{g288CayleyQuotient}. Finally, the largest $1152$-vertex reachability graph connects to $9$ isomorphic copies of itself under the action of $P_1$ and $P_2$. In the following section, we move beyond the set of stabilizer states and consider the action of $\langle H_1,\,H_2,\,C_{1,2},\,C_{2,1}\rangle$ on some notable non-stabilizer states.

\subsection{Non-Stabilizer Quotients}

Our state-independent description of the Clifford group allows us to examine the action of $\mathcal{C}_n$ on states which are not stabilizer states. For a quantum information theorist, the term ``stabilizer states'' typically refers to the set of $n$-qubit quantum states which are stabilized by a $2^n$-element subset of the Pauli group. There exist, however, states which are not stabilizer states, but are stabilized by additional Clifford group elements besides $\mathbb{1}$. These states likewise admit reachability graphs through our quotient procedure, and their graph properties reflect their distinction from the set of stabilizer states. Below we give a few examples of reachability graphs for notable non-stabilizer states, under the action of $\langle H_1,\,H_2,\,C_{1,2},\,C_{2,1}\rangle$, and contrast their structure with the stabilizer state graphs.

The $n$-qubit $W$-state holds particular interest as a highly-entangled, non-biseparable quantum state \cite{Dur2000,Schnitzer:2022exe}. Defined as
    \begin{equation}
        \ket{W}_n \equiv \left(\ket{100...00} + \ket{010...00} + ... + \ket{000...01} \right),
    \end{equation}
 $\ket{W}_n$ is famously not a stabilizer state when $n \geq 3$. However, $\ket{W}_n$ is stabilized by more than just $\mathbb{1}$ in $\mathcal{C}_n$. Even considering just the action of $\HC$, $\ket{W}_n$ is stabilized by the four elements
\begin{equation}\label{W3StabGroup}
    \{\mathbb{1},\, H_2C_{1,2}H_2,\, H_1C_{1,2}H_2C_{2,1},\, H_2C_{1,2}C_{2,1}C_{1,2}H_1\}.
\end{equation}

\begin{figure}[h]
        \centering    
        \begin{overpic}[width=11.5cm]{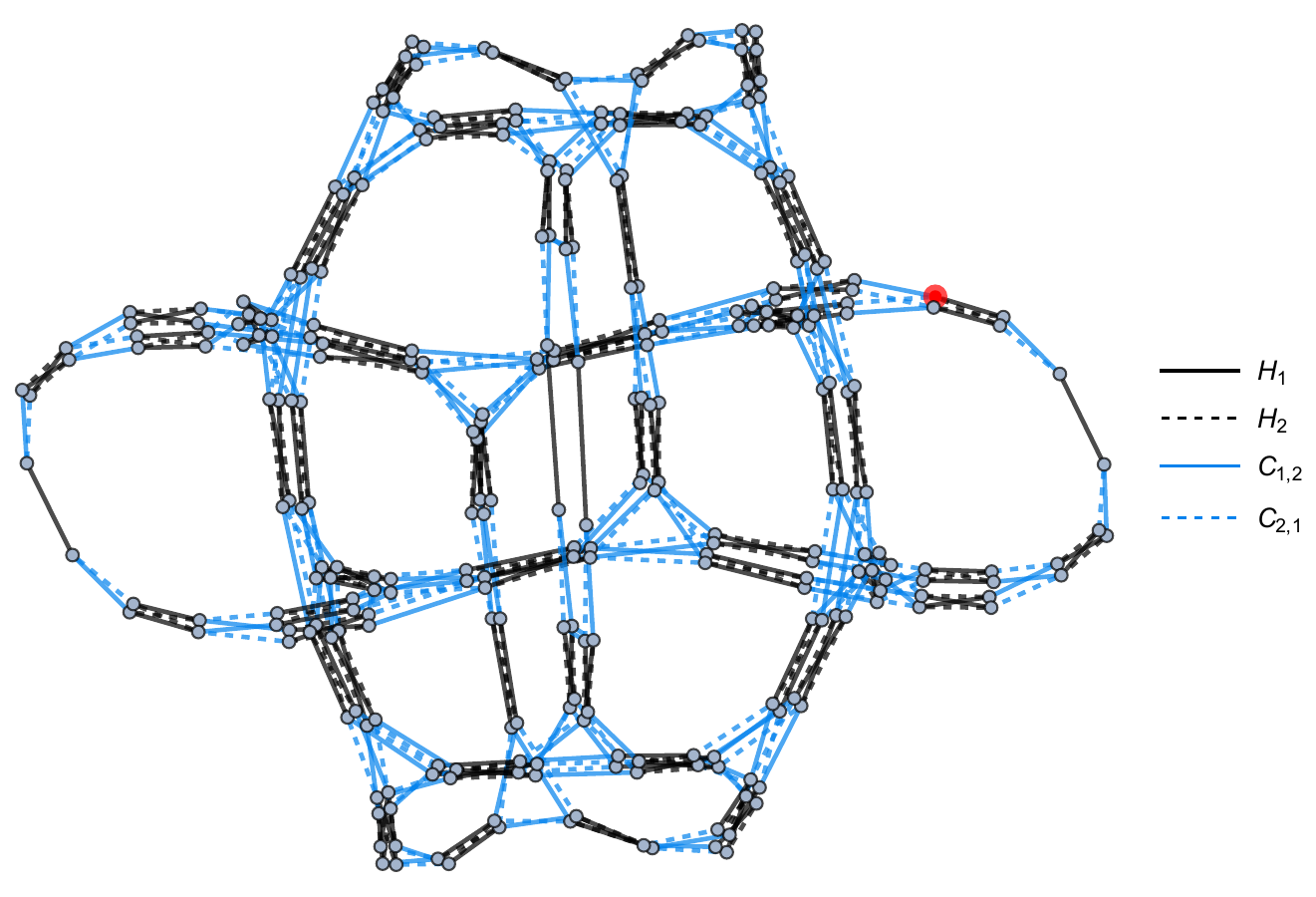}
            \put (72,47.5) {\footnotesize{$\swarrow \ket{W}_3$}}
        \end{overpic}
        \caption{Quotient space of $\HC$ Cayley graph after modding by $\ket{W}_n$ stabilizer subgroup. This reachability graph is not isomorphic to the $288$-vertex subgraph seen for stabilizer states in Figure \ref{g288CayleyQuotient}.}
    \label{D31HhCcGraph}
    \end{figure}

Figure \ref{D31HhCcGraph} shows the orbit of $\ket{W}_n$ under the action of $\HC$. The stabilizer subgroup of $\ket{W}_n$, in Eq. \eqref{W3StabGroup}, is isomorphic to all other stabilizer subgroups in the orbit seen in Figure \ref{D31HhCcGraph}. The stabilizer group of $\ket{W}_n$ is not, however, isomorphic to any subgroup in the orbit of the stabilizer state group in Eq. \eqref{g288StabGroup}. Consequently, while the reachability graph of $\ket{W}_n$ under $\HC$ contains $288$ vertices, its structure is distinctly different from the stabilizer state graph seen in Figure \ref{g288CayleyQuotient}.

Another notable set of non-stabilizer states are the $n$-qubit Dicke states \cite{Dicke1954,Baertschi2019}. Dicke states are equal superpositions of $n$-qubit basis states with Hamming weight $k$, defined
\begin{equation}
    \ket{D^n_k} \equiv \binom{n}{k}^{-1/2} \sum_{b \in \{0,1\}^n, \hspace{.1cm} h(b) = k} \ket{b},
\end{equation}
where $h(b)$ is the standard Hamming weight for binary stings. 

While $\ket{D^n_k}$ is not a stabilizer state%
\footnote{The state $\ket{D^n_n} = \ket{1}^{\otimes n}$ is a stabilizer state and its reachability graph is given in Figure \ref{g24CayleyQuotient}. Additionally, states $\ket{D^n_1} = \ket{W}_n$ and $\ket{D^n_{n-1}}$ have reachability graphs as shown in Figure \ref{D31HhCcGraph}.}
for all $n \neq k$ and $n >2$, every $\ket{D^n_k}$ is stabilized by more than $\mathbb{1}$ in $\HC$. States $\ket{D^n_k}$ where $1 < k < n-1$ are stabilized by exactly two elements of $\HC$, namely
\begin{equation}\label{D42StabGroup}
    \{\mathbb{1},\, H_2C_{1,2}C_{2,1}C_{1,2}H_1\}.
\end{equation}

\begin{figure}[h]
        \centering
        \begin{overpic}[width=10cm]{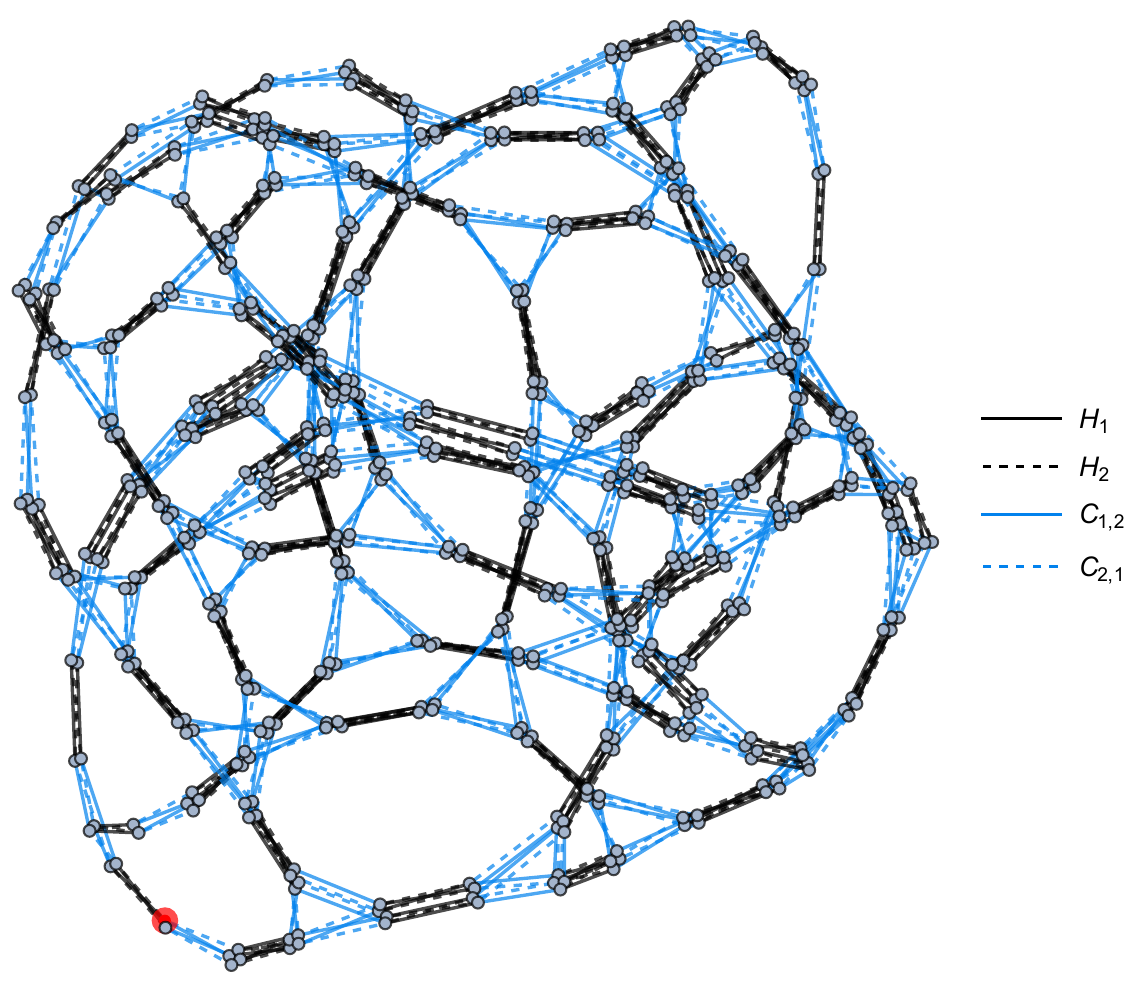}
            \put (1,5.7) {\footnotesize{$\ket{D^4_2}\rightarrow$}}
        \end{overpic}
        \caption{Quotient space of $\HC$ Cayley graph after modding by the stabilizer subgroup of $\ket{D^n_k}$, where $1 < k < n-1$. This reachability graph contains $576$ vertices, an orbit size not observed among the set of stabilizer states.}
    \label{D42HhCcGraph}
    \end{figure}

Figure \ref{D42HhCcGraph} shows an example reachability graph for the state $\ket{D^4_2}$. Since $\ket{D^4_2}$ is only stabilized by the two elements in Eq. \eqref{D42StabGroup}, its orbit under $\HC$ reaches $576$ states. Graphs with $576$ vertices are never observed among stabilizer states at any qubit number.

We have displayed the orbits of non-stabilizer states under the action of $\HC$, specifically for certain states which are stabilized by more than just the identity. We observed reachability graphs with vertex count not seen among the set of stabilizer states. Additionally, we identified graphs with vertex count shared with stabilizer states, but possessing a different topology. The orbits of Dicke states and their entanglement properties are studied in detail in \cite{Munizzi:2023ihc}.

\section{Discussion}\label{sec:discussion}

In this work we have presented a generalized construction for reachability graphs, defining them as quotient spaces of Cayley graphs. We began by constructing a presentation for $\mathcal{C}_1$ and $\mathcal{C}_2$, which we used to highlight the non-trivial relations among Clifford group elements. These relations allowed us to understand the structure of $\mathcal{C}_1$ and $\mathcal{C}_2$, and to explicitly construct all subgroups built from a restricted set of generators. Our operator-level, state-independent construction allowed us to obtain constraints on the evolution of any state through Clifford circuits. 

Extending our construction to higher qubit Clifford groups would require the definition of new relations  with each increase in qubit number. Intriguingly, extending our presentation for $\mathcal{C}_2$ to a presentation for $\mathcal{C}_3$ only requires the addition of $4$ relations \cite{Selinger2013}, none of which involve the phase gate.

After building their presentations, we studied Cayley graphs for $\mathcal{C}_1$ and $\mathcal{C}_2$, as well as all $\mathcal{C}_2$ subgroups generated by a subset of Hadamard, phase, and CNOT gates. Our protocol contracts the Cayley graph, yielding a quotient graph that is isomorphic to a state's reachability graph. Specifically, we quotient by the stabilizer subgroup for a particular state, ensuring only non-trivial action is represented by remaining edges. Using this procedure, we can analyze the evolution of a state through circuits comprised of the given gate set. Since we begin with the state-independent Cayley graph, this quotient protocol and analysis can be applied to any quantum state.

We emphasize that the techniques put forth in this paper are not limited to Clifford circuits. Any finite gate set can be represented by a discrete Cayley graph. The Cayley graph can then be made finite by imposing a cutoff on graph distance, which constrains the depth of a circuit. Accordingly, the program established in this paper could be used to study even universal gate sets in quantum computation, up to a fixed circuit depth. Our techniques could furthermore be straightforwardly extended to computation with qutrits or qudits.

Access to a graph-theoretic description of evolution through quantum circuits allows for direct calculation of some interesting circuit properties. For example, the gate complexity of a given circuit which transforms one state into another is precisely the minimum graph distance separating the two vertices that represent each state. The Cayley graph diameter hence immediately bounds the maximal change in complexity that can be observed under the constituent gate set. Additionally, given a fixed set of universal generators, one could compute complexity growth for circuits of varying depths. Conversely, one could fix a circuit depth and consider the growth of complexity under alternative sets of universal generators \cite{Munizzi_2022}. It would be interesting to relate the discrete picture of gate complexity obtained here to a more continuous picture, as in \cite{Balasubramanian:2019wgd}.

The graph analysis in this paper might be useful to better understand circuit architecture and reduce resource overhead in a quantum computation framework. The relations in our presentation often describe non-trivial, and sometimes unexpected, equivalences between sequences of quantum gates. In many cases, large-depth circuits containing strings of gates which are difficult to implement can be reduced to sequences of shorter depth, and simpler gate composition. One such example is $C_{i,j}H_jC_{i,j}P_jC_{i,j}P_j^3H_j = P_i$, where a circuit of $9$ gates, including numerous (and resource-expensive) $CNOT$ insertions, can be optimized to a single phase gate. Similarly, in the context of state preparation, an optimal circuit to transform an initial state into some desired final state can be identified by the appropriate extremal graph path. If computational or experimental constraints exist that limit the set of viable gates, corresponding edges in the graph can be modified or removed to accommodate this restriction. This analysis could be particularly interesting in the context of near-term quantum computing, where it is often easier to implement some specified set of gates than arbitrary two-body couplings.

In this work we focused on the group $\HC = \langle H_1,\,H_2,\,C_{1,2},\,C_{2,1} \rangle < \mathcal{C}_2$, which offered useful insight into the bipartite entanglement generated by Clifford circuits. Using our quotient procedure for the $\HC$ we were able to recover all stabilizer reachability graphs from our last paper, Figures \ref{g24CayleyQuotient}--\ref{g1152CayleyQuotient}, as well as reachability graphs for some non-stabilizer states. In particular, we showed that the $1152$-vertex graph is the largest reachability graph for any state, stabilizer or otherwise. We believe we have exhibited all reachability graphs involving $\mathcal{C}_2$ for stabilizer states, but proving so would require a deeper understanding of the relation between the Pauli stabilizer groups and Clifford stabilizer groups for a given state.

Instead of $\HC$, we could consider state orbits under alternative $\mathcal{C}_2$ subgroups. Orbits under different $\mathcal{C}_2$ subgroups should also decompose the full $\mathcal{C}_2$ reachability graph into disconnected pieces, similarly to the situation for $\HC$.

Additional stabilizer subgroups exist which quotient $\mathcal{C}_n$ and $\HC$ Cayley graphs, distinct from the stabilizer subgroups of individual stabilizer states. A $2$-element subgroup of $\HC$ stabilizes all Dicke states with a certain structure, building the reachability graph seen in Figure \ref{D42HhCcGraph}. Furthermore, all states we examined from which magic can be fault-tolerantly distilled \cite{bravyi2005universal,Veitch_2013} are stabilized by more than just $\mathbb{1}$ in $\mathcal{C}_n$. We conjecture that some measure of ``non-stabilizerness'', such as stabilizer Renyi entropy \cite{Leone:2021rzd} or mana \cite{White:2020zoz}, can be defined using the order of state's stabilizer subgroup in $\mathcal{C}_n$. 

We initiated this study to explore the evolution of entanglement entropy through Clifford circuits. Since entanglement in Clifford circuits can only be modified through $CNOT$ action, the number of $C_{i,j}$ edges in a reachability graph, which we term the ``CNOT diameter'', weakly bounds the number of times the entropy vector can change. However, our deeper exploration of the Clifford group revealed that not every $C_{i,j}$ gate modifies entropic structure, since relations like $(H_1C_{2,1})^4 = P_1^2$ demonstrate that some circuits with CNOT gates nonetheless never modify entanglement. In \cite{Keeler:2023shl}, we build ``contracted graphs'' which exhibit how entropy vectors can change within a given reachability graph.

\acknowledgments

The authors thank ChunJun Cao, Temple He, William Kretschmer, Daniel Liang, Julien Paupert, David Polleta, and Claire Zukowski for useful discussions.  CAK and WM are supported by the U.S. Department of Energy under grant number DE-SC0019470 and by the Heising-Simons Foundation ``Observational Signatures of Quantum Gravity'' collaboration grant 2021-2818. JP is supported by the Simons Foundation through
\emph{It from Qubit: Simons Collaboration on Quantum Fields, Gravity, and Information}.

\newpage

\begin{appendices}
\addtocontents{toc}{\protect\setcounter{tocdepth}{1}}

\section{$\mathcal{C}_2$ Table with Relations}\label{ExtendedTableAppendix}

Complete version of Table \ref{tab:OrbitLengthCliffordSubgroupNoRelations} with relations needed to generate each subgroup.

\begin{table}[h]
    \centering
    \begin{tabular}{|c||c|c||c|c||c|}
    \hline
    Generators & Order & Diam. & Fact. & Diam.* & Relation\\
    \hline
    \hline
    \{$H_1$\} & $2^\dagger$ & 1 & - & - & \ref{HSquared}\\
    \hline
    \{$C_{1,2}$\} & $2^\dagger$ & 1 & - & - & \ref{CSquared}\\
    \hline
    \{$P_1$\} & $4$ & 3 & - & - & \ref{PFourth}\\
    \hline
    \{$H_1,H_2$\} & $4$ & 2 & - & - & \ref{HSquared}, \ref{HhComm}\\
    \hline
    \{$C_{1,2},C_{2,1}$\} & $6$ & 3 & - & - & \ref{CSquared}, \ref{CcComm}\\
    \hline
    \{$H_1,P_2\}$ & $8^\dagger$ & 4 & - & - & \ref{HSquared}, \ref{PFourth}, \ref{HpComm}\\
    \hline
    \{$P_1,C_{1,2}\}$ & $8^\dagger$ & 4 & - & - & \ref{PFourth}, \ref{CSquared}, \ref{PCComm}\\
    \hline
    \{$P_1,P_2$\} & $16$ & 6 & - & - & \ref{PFourth}, \ref{PpComm}\\
    \hline
    \{$H_1,C_{2,1}\}$ & $16^\dagger$ & 8 & - & - & \ref{HSquared}, \ref{CSquared}, \ref{ChFourth}\\
    \hline
    \{$H_1,C_{1,2}\}$ & $16^\dagger$ & 8 & - & - & \ref{HSquared}, \ref{CSquared}, \ref{CNOTTransform}, \ref{ChFourth}\\
    \hline
    \{$H_1,P_2,C_{2,1}$\} & $32$ & 6 & - & - & \ref{HSquared}, \ref{PFourth}, \ref{CSquared}, \ref{HpComm}, \ref{PCComm}, \ref{ChFourth}\\
    \hline
    \{$P_1,C_{2,1}$\} & $32$ & 8 & - & - &  \ref{PFourth}, \ref{CSquared}, \ref{CpFourth}\\
    \hline
    \{$P_1,P_2,C_{2,1}$\} & $64$ & 7 & - & - &  \ref{PFourth}, \ref{CSquared}, \ref{PpComm}, \ref{CpFourth}, \ref{PCComm}\\
    \hline
    \{$P_1,C_{2,1},C_{1,2}$\} & $192$ & 11 & - & - & \ref{PFourth}, \ref{CSquared}, \ref{PpComm}, \ref{HpComm}--\ref{CNOTTransform}, \ref{CcComm}--\ref{ChFourth} \\
    \hline
    \{$H_1,P_1$\} & $192$ & 16 & 8 & 6 & \ref{HSquared}--\ref{HPComm}\\
    \hline
    \{$H_1,H_2,P_1$\} & $384$ & 17 & 8 & 7 & \ref{HSquared}--\ref{HPComm}, \ref{HhComm}, \ref{HpComm}\\
    \hline
    \{$P_1,P_2,H_1$\} & $768$ & 19 & 8 & 9 & \ref{HSquared}--\ref{HPComm}, \ref{HhComm}, \ref{HpComm}\\
    \hline
    \{$H_1,C_{2,1},C_{1,2}\}$ & $2304^*$ & 26 & 2 & 15 & \ref{HSquared}--\ref{HPComm}, \ref{CSquared}, \ref{CcComm}, \ref{ChFourth} \\
    \hline
    \{$H_1,H_2,C_{1,2}\}$ & $2304^*$ & 27 & 2 & 17 & \ref{HSquared}--\ref{HPComm}, \ref{CSquared}, \ref{CcComm}, \ref{ChFourth}\\
    \hline
    \{$H_1,H_2,C_{1,2},C_{2,1}\}$ & $2304^*$ & 25 & 2 & 15 & \ref{HSquared}--\ref{HPComm}, \ref{CSquared}, \ref{CcComm}, \ref{ChFourth}\\
    \hline
    \{$H_1,P_1,C_{2,1}\}$ & $3072^*$ & 19 & 8 & 9 & \ref{HSquared}--\ref{HPComm}, \ref{CSquared}, \ref{PpComm}, \ref{HpComm}, \ref{FourGenRelation}, \ref{CpFourth}, \ref{PCComm} \\
    \hline
    \{$H_1,P_1,C_{1,2}$\} & $3072$ & 19 & 8 & 11 & \ref{HSquared}--\ref{HPComm}, \ref{PpComm}, \ref{CpCpRelation}, \ref{CHpSquared}\\
    \hline
    \{$H_1,P_1,P_2,C_{2,1}\}$ & $3072^*$ & 19 & 8 & 9 & \ref{PFourth}, \ref{HPComm}, \ref{CSquared}, \ref{PpComm}, \ref{HpComm}--\ref{CNOTTransform}, \ref{CcComm}--\ref{ChFourth}\\
    \hline
    \{$H_1,H_2,P_1,P_2$\} & $4608$ & 17 & 8 & 12 & \ref{HSquared}--\ref{HPComm}, \ref{PpComm}, \ref{HpComm}\\
    \hline
    \{$H_1,P_2,C_{1,2}$\} & $9216$ & 24 & 8 & 13 & \ref{HSquared}, \ref{PFourth}, \ref{CSquared}, \ref{PpComm}, \ref{HpComm}--\ref{CNOTTransform}, \ref{PCComm}, \ref{ChFourth}\\
    \hline
    \{$H_1,H_2,P_1,C_{2,1}\}$ & $92160^*$ & 21 & 8 & 13 & \ref{HPComm}, \ref{PpComm}--\ref{CcComm}\\
    \hline
    \{$H_1,H_2,P_1,C_{1,2}\}$ & $92160^*$ & 21 & 8 & 16 & \ref{HPComm}, \ref{PpComm}--\ref{CcComm}\\
    \hline
    \{$H_1,P_1,P_2,C_{1,2}\}$ & $92160^*$ & 21 & 8 & 14 & \ref{HPComm}, \ref{PpComm}--\ref{CcComm}\\
    \hline
    All & $92160^*$ & 19 & 8 & 11 & \ref{HPComm}, \ref{PpComm}--\ref{CcComm}\\
    \hline
    \end{tabular}
\caption{$\mathcal{C}^2$ subgroups built by restricting generating set. Asterisk indicates two subgroups which contain the same elements, and a dagger indicates subgroups with isomorphic Cayley graphs. The relations needed to present each subgroup are given in the rightmost column.}
\label{tab:OrbitLengthCliffordSubgroup}
\end{table}

\newpage

\section{Derivation of Additional Relations}\label{AdditionalDerivationsAppendix}

In this Appendix, we provide a derivation of additional relations useful for subgroup construction, but not explicitly included in our presentation. Each of the relations below can be derived from Eqs. \eqref{HSquared}--\eqref{ChFourth}.

Using relations \ref{PFourth}, \ref{PpComm}, \ref{HpComm}, \ref{PCComm}, \ref{ChFourth}, \ref{FourGenRelation}, and \ref{CNOTTransform}, we can construct the useful identity,
\begin{equation}\label{CpCpRelationDerivation}
    C_{i,j}P_jC_{i,j}P_j = P_jC_{i,j}P_jC_{i,j}.
\end{equation}
A derivation of this identity is provided below for $C_{2,1}$ and $P_1$. For simplicity, let $H = H_1, h = H_2, C = C_{1,2},c = C_{2,1}, P=P_1,$ and $p=P_2$. We then derive,
    \begin{align*}
      HcH &= HcH, \nonumber \\
      HcHP &= HcHP, \nonumber \\
      hChP &= HcHP, \nonumber \\
      PhCh &= HcHP, \nonumber \\
      PHcH &= HcHP, \nonumber \\
      cPHcH &= cHcHP, \nonumber \\
      cPHcH &= p^2HcHcP, \nonumber \\
      p^2cPHcH &= HcHcP, \nonumber \\
      pcPHcpH &= HcHcP, \nonumber \\
      pcPccHcpH &= HcHcP, \nonumber \\  
      pcPcPcP^3 &= HcHcP, \nonumber \\  
      pcPcPcP^2 &= HcHc, \nonumber \\  
      cPcPcP^2p &= HcHc, \nonumber \\ 
      cPcPcP^2cHcpH &= \mathbb{1}, \nonumber \\
      ^* \qquad cPcPcP^3cP^3 &= \mathbb{1}, \nonumber \\
      cPcP &= PcPc, \nonumber
    \end{align*}
    where \ref{FourGenRelation} was used to show $cP^3cP^3 = cP^2cHcpH$, and acquire the starred line from the one above it.

A useful identity, derivable from relations \ref{PFourth}, \ref{CNOTTransform} and, \ref{ChFourth}, is the following,
\begin{equation}\label{CHPSquaredIdentity}
    (C_{i,j}H_iP_j^2)^2 = (P_j^2H_iC_{i,j})^2.
\end{equation}
A derivation of \eqref{CHPSquaredIdentity} is given below using $H_1, C_{1,2}$ and $P_2$. For simplicity, we again let $H = H_1, h = H_2, C = C_{1,2},c = C_{2,1},$ and $p=P_2$. We have,
    \begin{align*}
      p^4 = (p^2)^2 = (Ch)^8 &= \mathbb{1}, \nonumber \\
      ChChChChChChChCh &= \mathbb{1}, \nonumber \\
      CHcHCHcHCHcHCHcH &= \mathbb{1}, \nonumber \\
      (CH(cH)^4)^4 &= \mathbb{1}, \nonumber \\
      (CHp^2)^4 &= \mathbb{1}, \nonumber \\
      (C_{1,2}H_1P_2^2)^2 &= (P_2^2H_1C_{1,2})^2, \nonumber
    \end{align*}

A useful identity for constructing $\langle H_i, H_j, C_{i,j}\rangle$ is the following,
\begin{equation}\label{HTransformIdentity}
    C_{j,i}C_{i,j}C_{j,i}H_iC_{j,i}C_{i,j}C_{j,i} = H_j.
\end{equation}

A useful identity for constraining entropy in $\langle H_i, H_j, C_{i,j}\rangle$ reachability graphs is $(C_{i,j}H_j)^4 = P_i^2$. This relation can be proved using other relations in our presentation, beginning with the fact that $C_{i,j}$ and $P_i$ commute,
\begin{equation}\label{ChFourthEqualsPSquared}
\begin{split}
    C_{i,j}P_i&=P_iC_{i,j}, \\
    C_{i,j}P_iH_j&=P_iC_{i,j}H_j, \\
    C_{i,j}H_jP_i&=P_iC_{i,j}H_j, \\
    C_{i,j}H_jC_{i,j}P_iC_{i,j}&=C_{i,j}P_iH_j, \\
    C_{i,j}H_jC_{i,j}P_i&=C_{i,j}P_iH_jC_{i,j}, \\
    P_jC_{i,j}P_j^3H_j&=H_jP_j^3C_{i,j}P_j, \\
    C_{i,j}P_j^3H_jP_j^3&=P_j^3H_jP_j^3C_{i,j}, \\
    H_j&=P_jC_{i,j}P_j^3H_jP_j^3C_{i,j}P_j, \\
    H_j&=P_jC_{i,j}P_j^3C_{i,j}P_iH_jC_{i,j}, \\
C_{i,j}H_jC_{i,j}H_j&=C_{i,j}H_jC_{i,j}P_jC_{i,j}P_j^3H_jH_jC_{i,j}P_iH_jC_{i,j}, \\
C_{i,j}H_jC_{i,j}H_j&=P_iH_jC_{i,j}P_iH_jC_{i,j}, \\
(C_{i,j}H_j)^4 &= P_i^2.\\
\end{split}
\end{equation}

\section{Detailed Subgroup Presentations}\label{SubgroupPresentationAppendix}

The details of building $\langle H_1,\,P_2,\,C_{1,2} \rangle$ are given below.
        \begin{enumerate}
        \item For words containing $0$ $H_1$ operations, we have the $32$ elements $b \in \langle C_{1,2},\,P_2 \rangle = \{p,\, pC_{1,2}p,\, C_{1,2}\overline{p}C_{1,2}p\}$, with $p$ defined as in Eq. \eqref{PhaseGroup}.
        \item Words containing $1$ $H_1$ have the form $bH_1b$. Since $H_1$ and $P_2$ commute, we can push all $P_2$ operations to the right until they reach a $C_{1,2}$. This action corresponds to multiplying all $0$ $H_1$ words on the left by $H_1,\, pC_{1,2}H_1,$ and $C_{1,2}\overline{p}C_{1,2}H_1$, giving the set $\{H_1b,\, pC_{1,2}H_1b,\, C_{1,2}\overline{p}C_{1,2}H_1b\}$, which has $32 + (4 \times 32) + (3 \times 32) = 256$ elements. 
         \item To generate words with $2$ $H_1$ operations, we left-multiply all $1$ $H_1$ words that do not begin with $H_1$, by the set $\{H_1, pC_{1,2}H_1,C_{1,2}\overline{p}C_{1,2}H_1\}$ (since otherwise we would collapse resulting $H_1$ pair). This procedure generates $1792$ elements, with $512$ duplicates such as
         \begin{equation}
                H_1C_{1,2}H_1 = P_2^2C_{1,2}H_1P_2^2C_{1,2}H_1P_2^2C_{1,2}P_2^2,
         \end{equation}
         as well as, 
         \begin{equation}
                C_{1,2}H_1C_{1,2}H_1 = C_{1,2}P_2^2C_{1,2}H_1P_2^2C_{1,2}H_1P_2^2C_{1,2}P_2^2,
         \end{equation}
         both reducible by Eq. \eqref{CHpSquared}. After removing duplicates, there are $1280$ unique $2$ $H_1$ words added to our set.

         \item For words containing $3$ $H_1$ operations, we again left-multiply all $2$ $H_1$ words which \underline{do not} begin with $H_1$, by $\{H_1, pC_{1,2}H_1,$ and $C_{1,2}\overline{p}C_{1,2}H_1\}$, generating $12544$ elements. 

         Of these $12544$ elements, $9152$ are duplicates of other $3$ $H_1$ words, e.g.
         \begin{equation}
                    H_1C_{1,2}H_1C_{1,2}H_1 = H_1C_{1,2}P_2^2C_{1,2}H_1P_2^2C_{1,2}H_1P_2^2C_{1,2}P_2^2,
         \end{equation}
         described by Eq. \eqref{CHpSquared}. An additional $256$ are duplicates of $1$ $H_1$ words, e.g.
         \begin{equation}
                H_1C_{1,2}H_1P_2^2C_{1,2}H_1 = P_2^2C_{1,2}H_1P_2^2C_{1,2}P_2^2,
         \end{equation}
         also described by Eq. \eqref{CHpSquared}, leaving only $3136$ new contributions to the subgroup.
        
         \item Words with $4$ $H_1$ operations are likewise built by left-multiplying all $3$ $H_1$ words that \underline{do not} begin with $H_1$, by $\{H_1, pC_{1,2}H_1,$ and $C_{1,2}\overline{p}C_{1,2}H_1\}$. This process generates $87808$ elements, $83200$ of which are duplicates of other $4$ $H_1$ words, e.g.
                \begin{equation}
                    H_1C_{1,2}H_1C_{1,2}H_1C_{1,2}H_1 = C_{1,2}H_1C_{1,2}H_1C_{1,2}H_1C_{1,2}H_1C_{1,2},
                \end{equation}
         which can be reduced by $(H_1C_{1,2})^8 = \mathbb{1}$. Another $1280$ are duplicates of $2$ $H_1$ words, such as
         \begin{equation}
             H_1C_{1,2}H_1C_{1,2}H_1P_2^2C_{1,2}H_1 = H_1C_{1,2}P_2^2C_{1,2}H_1P_2^2C_{1,2}P_2^2,
         \end{equation}
         described by Eqs. \eqref{CHpSquared} and \eqref{CpCpRelation}. A final $32$ elements are duplicates of $0$ $H_1$ words, e.g.
         \begin{equation}
             H_1C_{1,2}H_1P_2^2C_{1,2}H_1P_2^2C_{1,2}H_1 = P_2^2C_{1,2}P_2^2,
         \end{equation}
         explained using Eq. \eqref{CHpSquared}. Removing duplicates adds $3296$ new $4$ $H_1$ words to our subgroup.
        
         \item Finally we construct words with $5$ $H_1$ operations, multiplying all $4$ $H_1$ words that \underline{do not} begin with $H_1$ by $H_1, pC_{1,2}H_1,$ and $C_{1,2}\overline{p}C_{1,2}H_1$, and generating $614656$ words. Of these $614656$ words, $609792$ are duplicates of other $5$ $H_1$ words, e.g.
         \begin{equation}
         \begin{split}
         H_1P_2C_{1,2}&H_1P_2^3C_{1,2}H_1P_2^3C_{1,2}H_1P_2C_{1,2}H_1P_2C_{1,2}P_2^3\\
        &= C_{1,2}H_1C_{1,2}H_1C_{1,2}H_1C_{1,2}H_1C_{1,2}H_1,
         \end{split}
         \end{equation}
         described by Eq. \eqref{CHpSquared} using relations \ref{PFourth}, \ref{CSquared}, and \ref{CpFourth}. Another $3392$ are duplicates of $3$ $H_1$ words, e.g.
            \begin{equation}
                    C_{1,2}H_1C_{1,2}H_1C_{1,2}H_1C_{1,2} = H_1C_{1,2}H_1C_{1,2}H_1C_{1,2}H_1C_{1,2}H_1,
            \end{equation}
         described by $(H_1C_{1,2})^8 = \mathbb{1}$, and a final $256$ are duplicates of $1$ $H_1$ words, e.g.
                \begin{equation}
                    H_1 = H_1C_{1,2}H_1C_{1,2}H_1P_2^2C_{1,2}H_1P_2^2C_{1,2}H_1C_{1,2}P_2^2C_{1,2}P_2^2,
                \end{equation}
         by Eqs. \eqref{CHpSquared} and \eqref{CpCpRelation}. Upon removing duplicates, there are $1216$ unique $5$ $H_1$ words added to our subgroup, giving an order of $9216$.
        \end{enumerate}

\section{Magma Algebra System}\label{MagmaAppendix}

The Magma Computer Algebra system \cite{MR1484478} was used to validate many of the subgroup constructions in this paper. This program, which can be downloaded to a local device or accessed via a browser at \href{http://magma.maths.usyd.edu.au/magma/}{http://magma.maths.usyd.edu.au/magma/}, offers a powerful suite of group-theoretic calculators. We have included below a few examples of code input and output that demonstrate some functionality, and provide the interested reader a coarse template for using the software.

We formally construct subgroups of $\mathcal{C}_1$ and $\mathcal{C}_2$ by taking quotient groups of the free group generated by a select set of operations. For example, the group $\langle H_i,\,P_i \rangle$ can be built as
\inBox{$F\langle H,P \rangle := \textnormal{FreeGroup}(2);$\\
$G\langle x, y\rangle, \textnormal{phi} := \textnormal{quo}\langle F|H^2 = P^4 = 1, (H*P)^3 = (P*H)^3\rangle;$\\
$G$}
\outBox{Finitely presented group $G$ on $2$ generators\\
Relations\\
\begin{equation*}
    \begin{split}
    x^2 &= \textnormal{Id}(G)\\
    y^4& = \textnormal{Id}(G)\\
    (x * y)^3 &= (y * x)^3\\
    \end{split}
\end{equation*}}

Further calculations can be used to yield desired group properties such as order, number of defining generators, checks for abelianess, cyclicity, and many more. Some simple examples are included below.
\inBox{$F\langle P,C \rangle := \textnormal{FreeGroup}(2);$\\
$G\langle x, y\rangle, \textnormal{phi} := \textnormal{quo}\langle F|C^2 = P^4 = 1, C*P*C=P\rangle;$\\
Order(G)}
\outBox{$8$}
\inBox{$F\langle H,c \rangle := \textnormal{FreeGroup}(2);$\\
$G\langle x, y\rangle, \textnormal{phi} := \textnormal{quo}\langle F|c^2 = H^2 = (H*c)^8 = 1;$\\
Parent(G)}
\outBox{Power Structure of GrpFP}

To generate the complete two-qubit Clifford group $\mathcal{C}_2$, we include all generators and the relations of our presentation. We can subsequently simplify the presentation with the following code.
\inBox{$F<H,\,h,\,C,\,c,\,P,\,p> := \textnormal{FreeGroup}(6);$\\
$G<u,\,v,\,w,\,x,\,y,\,z>, \textnormal{phi} := \textnormal{quo}\langle F|  P^4 = p^4 = H^2 = h^2 = C^2 = c^2 = (H*C*p^2)^4 = (c*C)^3  =(c*H)^8 = 1,\,
H*h*H = h,\, H*p*H = p,\, h*P*h = P,\, P^3*p*P = p,\, C*P*C = P,\, c*p*c = p,\, (H*P)^3 = (P*H)^3,\, (p*h)^3 = (h*p)^3,\,  P*H*P = (c*p*h)^3,\, C*h*C*p*C*(p^3)*h = P,\, c*C*c*H*c*C*c = h,\, H*h*c*H*h = C,\, c*C*p^2*C*c = P^2,\,  p*C*p*C = C*p*C*p,\, P^2 = (C*h)^4,\, P^2 = (C*p)^4>;$\\
Order(G);\\
Simplify(G)}
\outBox{$92160$\\
Finitely presented group on 4 generators\\
Generators as words in group G\\
\begin{equation*}
    \begin{split}
    \$.1 &= u\\
    \$.2 &= v\\
    \$.3 &= w\\
    \$.4 &= z\\
    \end{split}
\end{equation*}
Relations\\
  \$.3$\wedge$2 = Id(\$)\\
  \$.1$\wedge$2 = Id(\$)\\
  \$.2$\wedge$2 = Id(\$)\\
  (\$.1 * \$.2)$\wedge$2 = Id(\$)\\
  \$.4$\wedge$4 = Id(\$)\\
  \$.1 * \$.4 * \$.1 * \$.4$\wedge$-1 = Id(\$)\\
   \$.2 * \$.4$\wedge$-1 * \$.3 * \$.4 * \$.2 * \$.4 * \$.3 * \$.4$\wedge$-1 = Id(\$)\\
   \$.2 * \$.3 * \$.2 * \$.4$\wedge$-1 * \$.2 * \$.3 * \$.2 * \$.4 = Id(\$)\\
   \$.4 * \$.3 * \$.4 * \$.3 * \$.4$\wedge$-1 * \$.3 * \$.4$\wedge$-1 * \$.3 = Id(\$)\\
   (\$.4$\wedge$-1 * \$.3 * \$.4$\wedge$2 * \$.3 * \$.4$\wedge$-1)$\wedge$2 = Id(\$)\\
   \$.2 * \$.4 * \$.3 * \$.4$\wedge$-1 * \$.2 * \$.3 * \$.2 * \$.4$\wedge$-1 * \$.3 * \$.4 * \$.2* \$.3 = Id(\$)\\
 \$.3 * \$.4$\wedge$-1 * \$.3 * \$.2 * \$.3 * \$.2 * \$.3 * \$.4 * \$.3 * \$.2 * \$.3 * \$.2 = Id(\$)\\
 \$.4 * \$.2 * \$.4 * \$.2 * \$.4 * \$.2 * \$.4$\wedge$-1 * \$.2 * \$.4$\wedge$-1 * \$.2 * \$.4$\wedge$-1 * \$.2 = Id(\$)\\
  \$.2 * \$.1 * \$.3 * \$.2 * \$.1 * \$.4 * \$.2 * \$.1 * \$.3 * \$.2 * \$.1 * \$.4$\wedge$-1 =Id(\$)\\
  \$.1 * \$.3 * \$.4$\wedge$2 * \$.1 * \$.3 * \$.4$\wedge$-2 * \$.1 * \$.3 * \$.4$\wedge$-2 * \$.1 * \$.3 *
    \$.4$\wedge$-2 = Id(\$)\\
    \$.3 * \$.2 * \$.1 * \$.3 * \$.2 * \$.1 * \$.3 * \$.1 * \$.3 * \$.2 * \$.1 * \$.3 * \$.2
    * \$.1 * \$.3 * \$.2 = Id(\$)\\
    (\$.3 * \$.1)$\wedge$8 = Id(\$)\\
    \$.4$\wedge$-1 * \$.1 * \$.3 * \$.2 * \$.1 * \$.3 * \$.4$\wedge$-2 * \$.3 * \$.4 * \$.1 * \$.2 * \$.3*\\ \$.1 * \$.2 * \$.3 * \$.4$\wedge$-1 * \$.2 * \$.4 * \$.3 = Id(\$)\\
    \$.2 * \$.3 * \$.2 * \$.4$\wedge$-1 * \$.3 * \$.1 * \$.3 * \$.4 * \$.3 * \$.2 * \$.3 * \$.2*\\
    \$.4$\wedge$-1 * \$.1 * \$.2 * \$.3 * \$.4$\wedge$-1 * \$.2\\ * \$.3 * \$.4$\wedge$-1 * \$.2 * \$.3 * \$.1 *\$.3 = Id(\$)\\
    \$.3 * \$.4 * \$.3 * \$.2 * \$.3 * \$.1 * \$.4 * \$.3 * \$.4$\wedge$-1 * \$.2 * \$.3 * \$.2\\
    *\$.3 * \$.1 * \$.4 * \$.3 * \$.4$\wedge$-1 * \$.2 * \$.3 * \$.2 * \$.3 * \$.1 * \$.3 * \$.2
    *\$.3 * \$.4$\wedge$-1 * \$.3 * \$.4 * \$.1 * \$.2 * \$.3 * \$.2 * \$.3 * \$.4$\wedge$-1 * \$.3 * \$.4* \$.1 * \$.2 * \$.3 * \$.2 * \$.3 * \$.4$\wedge$-1 * \$.3 * \$.1 = Id(\$)}

\section{Additional Graphs}\label{AdditionalGraphsAppendix}

In this section we include several additional graphs, not included in the main text. Each figure below offers further visualization for relations \eqref{HSquared}--\eqref{ChFourth}.

\begin{figure}[h]
        \centering        
        \includegraphics[width=8.5cm]{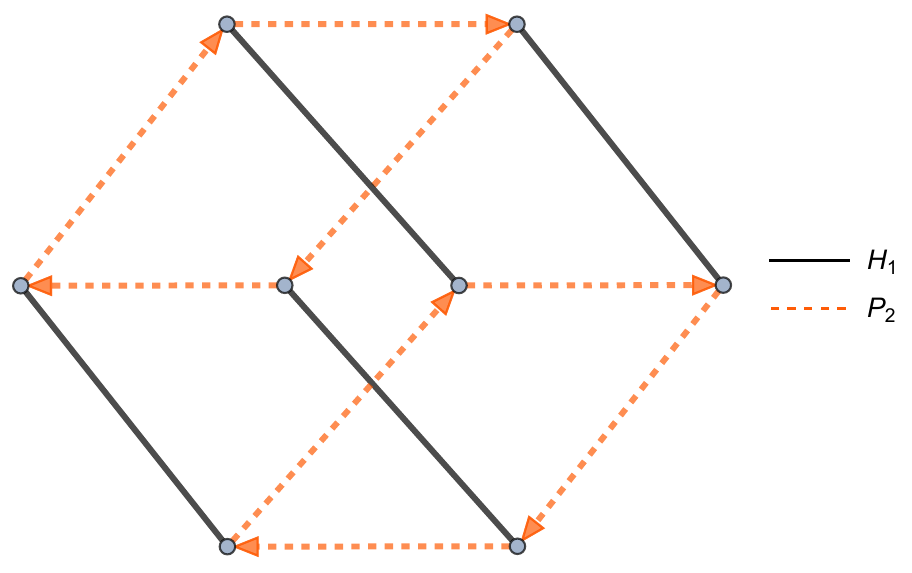}
        \caption{Cayley graph for $\langle H_1,\,P_2 \rangle$, the direct product $\langle H_1 \rangle \times \langle P_2 \rangle$, with $8$ vertices. Individual group structures for $\langle H_1 \rangle$ and $\langle P_2 \rangle$ are easily verified.}
        \label{H1P2CayleyGraph}
    \end{figure}

Figure \ref{H1P2CayleyGraph} shows the Cayley graph for $\langle H_1,\,P_2 \rangle$, containing $8$ vertices. Since $H_1$ and $P_2$ commute, the group $\langle H_1,\,P_2 \rangle$ is the direct product of $\langle H_1 \rangle \times \langle P_2 \rangle$, which is manifest in the Cayley graph structure.

\begin{figure}[h]
        \centering        
        \includegraphics[width=8.8cm]{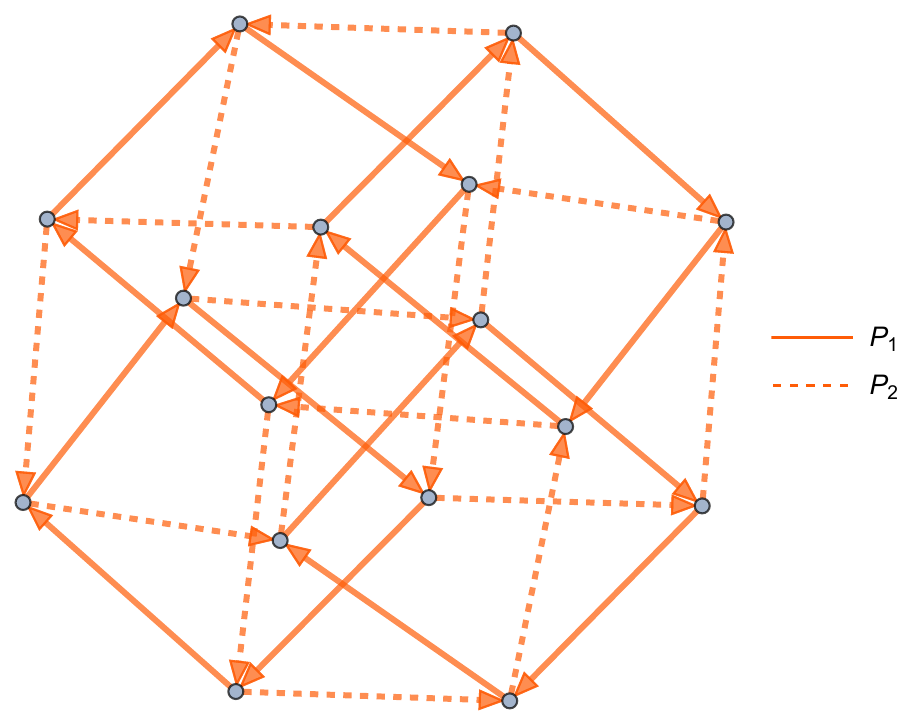}
        \caption{Cayley graph of $\langle P_1,\,P_2 \rangle$, the direct product $\langle P_1 \rangle \times \langle P_2 \rangle$, containing $16$ vertices.}
        \label{P1P2CayleyGraph}
    \end{figure}

\begin{figure}[h]
        \centering        
        \includegraphics[width=9cm]{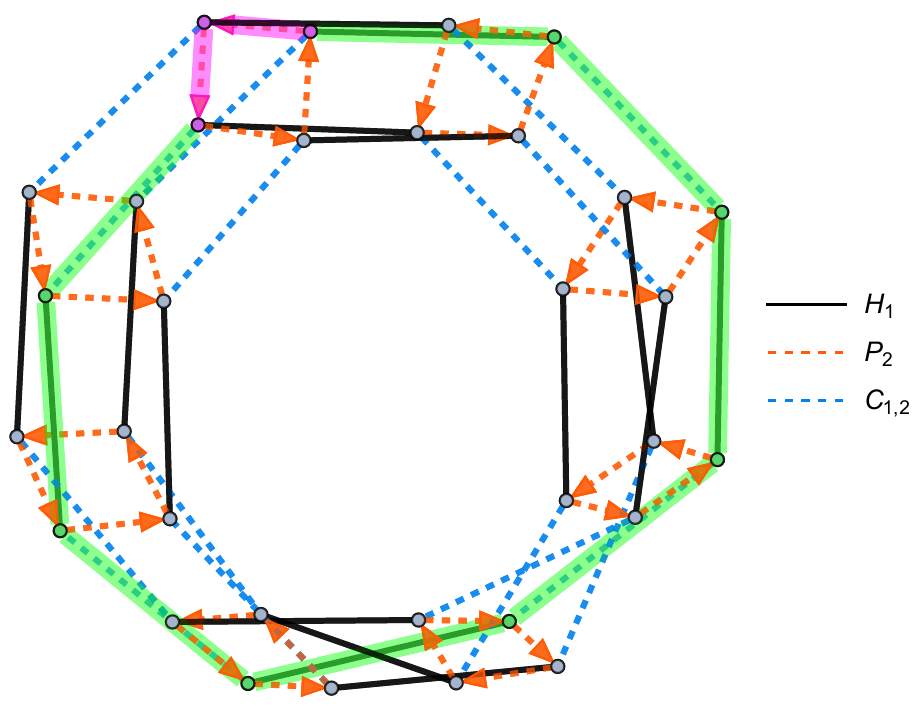}
        \caption{Cayley graph of $\langle H_1,\,P_2,\,C_{1,2} \rangle$, with $32$ vertices, where we note the non-trivial relation $(H_1C_{1,2})^4 = P_2^2$. Sequence $(H_1C_{1,2})^4$ is highlighted in green, and $P_2^2$ in magenta.}
        \label{H1P2C12CayleyGraph}
    \end{figure}

Figure \ref{P1P2CayleyGraph} shows the Cayley graph of $\langle P_1,\,P_2 \rangle$. The graph has $16$ vertices, and corresponds to the direct product $\langle P_1 \rangle \times \langle P_2 \rangle$.
\newpage
The group $\langle H_1,\,P_2,\,C_{1,2} \rangle$ is represented by the Cayley graph in Figure \ref{H1P2C12CayleyGraph}. The graph contains $32$ vertices, and the relation $(H_1C_{1,2})^4 = P_2^2$ can be directly visualized.

\begin{figure}[h]
        \centering        
        \includegraphics[width=9.5cm]{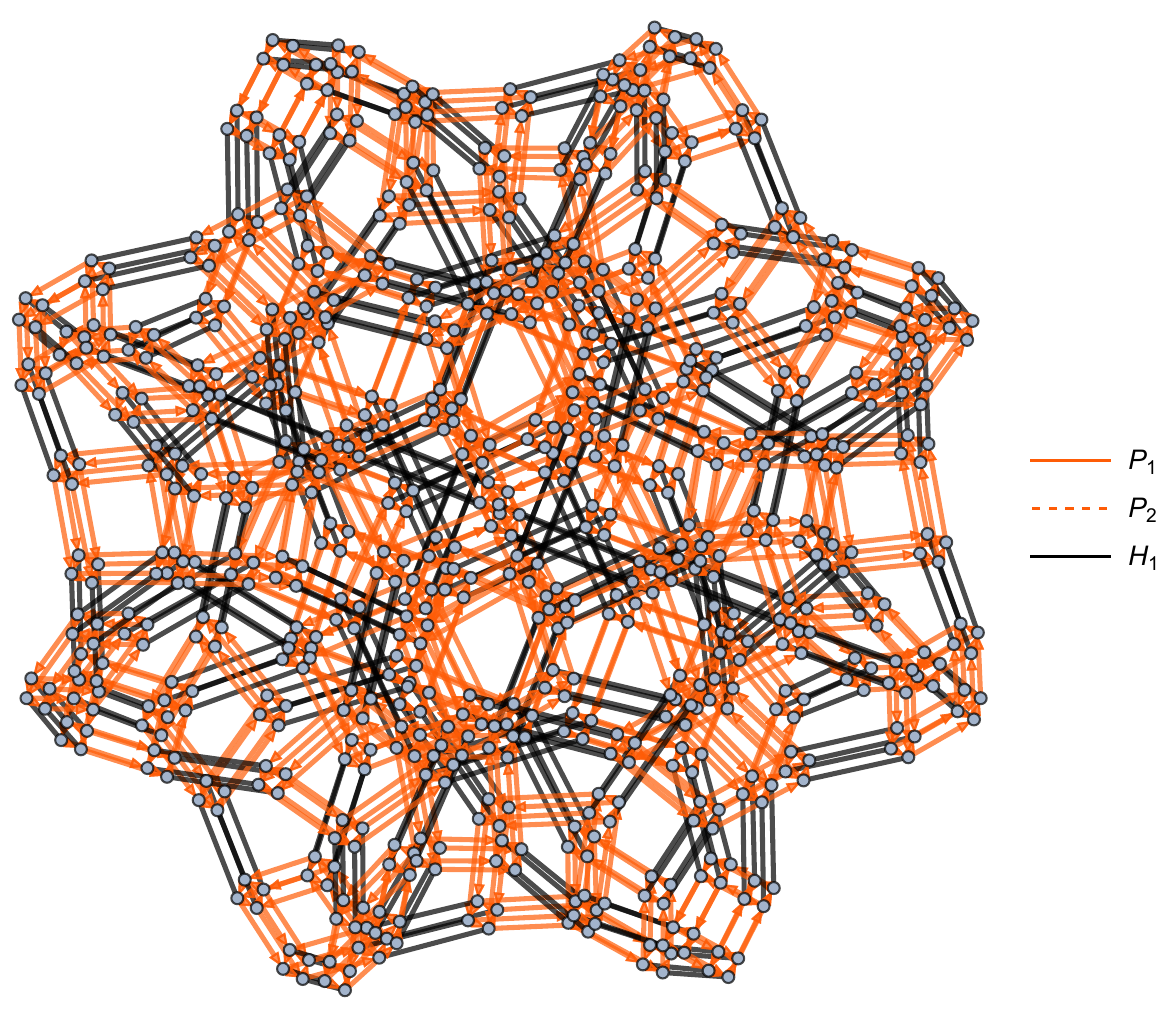}
        \caption{Cayley graph of $\langle P_1,\,P_2,\,H_1 \rangle$ with $768$ vertices. The graph contains directed $P_1$ and $P_2$ boxes connected by $H_1$ edges.}
        \label{P1P2H1CayleyGraph}
    \end{figure}

Figure \ref{P1P2H1CayleyGraph} gives the Cayley graph for $\langle P_1,\,P_2,\,H_1 \rangle$. While the graph is large, the symmetric structure of stacked $P_1$ and $P_2$ boxes connected by $H_1$ can be observed.

\end{appendices}

\bibliographystyle{JHEP}
\bibliography{CliffordGroupPaper}

\end{document}